# A Simultaneous Synergistic Protection Mechanism in Hybrid Perovskite-Organic Multi-junctions Enables Long-Term Stable and Efficient Tandem Solar Cells

Chao Liu[1,2,#]*, Kaicheng Zhang[2,#]*, Xin Zhou[4], Mingjian Wu[4], Paul Weitz[2], Shudi Qiu[2], Andrej Vincze[5], Yuchen Bai[2], Michael A. Anderson[6,7], Johannes Frisch[6,7], Regan G. Wilks[6,7], Marcus Bär[1,6,7,12], Zijian Peng[2,9], Chaohui Li[2,9], Jingjing Tian[2,9], Jiyun Zhang[2], Jianchang Wu[1], Jonas Englhard[8], Thomas Heumüller[1,2], Jens Hauch[1], Yixing Huang[10], Ning Li[1,11], Julien Bachmann[8], Erdmann Spiecker[4], and Christoph J. Brabec[1,2,3]*

[1]Helmholtz-Institute Erlangen-Nürnberg for Renewable Energy (HI ERN), Immerwahrstraße 2, 91058 Erlangen, Germany

[2]Institute of Materials for Electronics and Energy Technology (i-MEET), Friedrich-Alexander-Universität Erlangen-Nürnberg (FAU), Martensstr. 7, 91058 Erlangen, Germany

[3]Institute of Energy Materials and Devices (IMD-3), Forschungszentrum Jülich GmbH, Wilhelm-Johnen-Straße 52428 Jülich, Germany

[4]Institute of Micro- and Nanostructure Research & Center for Nanoanalysis and Electron Microscopy (CENEM), Friedrich-Alexander-Universität Erlangen-Nürnberg (FAU), IZNF, Cauerstr. 3, 91058 Erlangen, Germany

[5]International Laser Centre SCSTI, 84104 Bratislava, Slovak Republic

[6]Department Interface Design, Helmholtz-Zentrum Berlin für Materialien und Energie GmbH (HZB), Albert-Einstein-Str. 15, 12489 Berlin, Germany

[7]Energy Materials In-situ Laboratory Berlin (EMIL), Helmholtz-Zentrum Berlin für Materialien und Energie GmbH (HZB), Albert-Einstein-Str. 15, 12489 Berlin, Germany

[8]Chemistry of Thin Film Materials, Department of Chemistry and Pharmacy, Friedrich-Alexander-Universität Erlangen-Nürnberg (FAU), IZNF, Cauerstr. 3, 91058 Erlangen, Germany

[9]Erlangen Graduate School in Advanced Optical Technologies (SAOT), Paul-Gordan-Straße 6, 91052 Erlangen, Germany

[10]Institute of Medical Technology, Health Science Center, Peking University, Xueyuan Rd. 38, Haidian District, 100083, Beijing, China

[11]Institute of Polymer Optoelectronic Materials & Devices, Guangdong Basic Research Center of Excellence for Energy & Information Polymer Materials, State Key Laboratory of Luminescent Materials & Devices, South China University of Technology, Guangzhou, 510640, P. R. China

[12]Department of Chemistry and Pharmacy, Friedrich-Alexander-Universität Erlangen-Nürnberg (FAU), Egerlandstr. 3, 91058 Erlangen, Germany

[#]These authors contributed equally: Chao Liu, Kaicheng Zhang

*Correspondence:

c.liu@fz-juelich.de (C.L.),

kaichengzhang@zoho.com (K.Z.),

christoph.brabec@fau.de (C.B.)

## ABSTRACT

Perovskite-organic tandem solar cells (P-O TSCs) hold great promise for next-generation thin-film photovoltaics, with steadily improving power conversion efficiency (PCE). However, the development of optimal interconnecting layers (ICLs) remains one major challenge for further efficiency gains, and progress in understanding the improved long-term stability of P-O tandem configuration has been lagging. In this study, we experimentally investigate the enhanced stability of p-i-n P-O TSCs employing a simplified $C_{60}$/atomic-layer-deposition (ALD) $SnO_x$/PEDOT: PSS ICL without an additional charge recombination layer (CRL), which achieve an averaged efficiency of 25.12% and a hero efficiency of 25.5%. Our finding discovers that the recrystallization of $C_{60}$, a widely used electron transport layer in perovskite photovoltaics, leads to the formation of grain boundaries during operation, which act as migration channels for the interdiffusion of halide and Ag ions. Critically, we demonstrate for the first time that the tandem device architecture, incorporating organic semiconductor layers, effectively suppresses the bi-directional ion diffusion and mitigates electrode corrosion. Thus, the P-O TSC establishes a mutual protection system: the organic layers stabilize the perovskite sub-cell by suppressing ion diffusion-induced degradation, and the perovskite layer shields the organic sub-cell from spectrally induced degradation. The simultaneous synergistic protection mechanism enables P-O TSCs to achieve exceptional long-term operational stability, retaining over 91% of their initial efficiency after 1000 hours of continuous metal-halide lamp illumination, and to exhibit minimal fatigue after 86 cycles (2067 hours) of long-term diurnal (12/12-hour) testing. These results demonstrate that tandem cells significantly outperform their single-junction counterparts in both efficiency and stability.

## INTRODUCTION

Tandem solar cells (TSCs), integrating both wide and narrow bandgap (NBG) photo absorbers with complementary absorption spectra, have ignited tremendous interest, offering the potential to surpass the Shockley-Queisser (S-Q) limit[1-3]. Halide perovskite materials, featured with a broad bandgap ($E_g$) tunability, render them excellent building blocks as both front and rear sub-cells in TSCs[4,5]. Recently, perovskite-based TSCs have reached certified power conversion efficiencies (PCEs) of 34.6% and 30.1% for perovskite-silicon TSCs and monolithic all-perovskite TSCs, respectively[6,7], much higher than the current record PCE of single-junction perovskite devices (26.95%)[6]. On the downside, the silicon cells are usually fabricated at high temperatures exceeding 400 °C, while the NBG Sn-based perovskites ($E_g$ < 1.4 eV) in all-perovskite tandem devices encounter the intractable oxidation of $Sn^{2+}$ to $Sn^{4+}$, triggering undesirable heavy p-type doping or device degradation[8-10]. Alternatively, the advent of non-fullerene acceptors (NFAs), featuring near-infrared-region absorption, non-toxicity, and orthogonal solution processibility at low-temperatures as well as a simple fabrication process, has ignited the vigorous performance progress of organic solar cells (OSCs) over 20%, which positions them as viable rear NBG sub-cell candidates for perovskite-based TSCs[11,12]. Accordingly, perovskite-organic TSCs (P-O TSCs), which combine organic and halide perovskite semiconductors, are expected to yield high open-circuit voltage ($V_{OC}$), and to optimize sunlight utilization, especially anticipating new generations of high performing low bandgap NFAs with Eg < 1.2 eV. The efficiencies of P-O TSCs have been improved rapidly, now achieving 26.7% (certified 26.4%)[13]. Despite these achievements, developing efficient monolithic P-O TSCs still faces significant challenges, particularly in further improving both efficiency and stability. Progress in these areas hinges on meticulous fabrication of each component, including high-quality wide bandgap (WBG) halide perovskite front sub-cells, robust and efficient interconnecting layers (ICLs), and optimized organic rear sub-cells with finely tuned morphologies and extended absorption in the near-infrared (NIR) range.[14-19] The ICLs, which electrically stack front and rear sub-cells in series, play a pivotal role in determining the overall performance of TSCs by influencing internal electrical contact and energy alignment, managing light utilization in rear sub-cells, and governing charge collection and recombination between the sub-cells. Therefore, the design of ideal ICLs must ensure good quasi-ohmic contact, efficient internal charge recombination, minimized electrical and optical losses, as well as low manufacturing cost. In state-of-the-art p-i-n P-O TSCs, ICLs are commonly composed of tin oxide ($SnO_x$) deposited by atomic-layer deposition (ALD) as an electron transporting layer (ETL) and an evaporated thin metal layer (Au or Ag) or a sputtered transparent metal oxide (indium zinc oxide[15] or indium tin oxide[20]) as a charge recombination layer (CRL). The incorporation of an additional CRL serves two critical functions: (1) compensating for insufficient electrical contact between ETL and hole transporting layer to achieve quasi-ohmic behaviour, and (2) offering sufficient charge recombination sites[14,17,21,22]. Unfortunately, the inclusion of thin metal layers as CRLs inevitably leads to light absorption loss in the rear OSC sub-cell due to the metal films' substantial parasitic localized surface plasma resonance absorption and light reflection[15,18,23], which in turn limits the full advantage of tandem structures. Additionally, metal elements can react with mobile halide species, which can contribute to the deterioration of device stability.[24-26] While highly transparent sputtered metal-oxide layers offer alternatives, the sputtering process risks damaging the underlying layers due to the high-energy ion generated in the sputter proces[27,28]. Furthermore, the incorporation of extra CRLs produced through different methods into prevalent ICLs complicates device fabrication processes and increases manufacturing costs, confining their practical application potential[14,15,18]. These considerations drive the exploration of CRL-free ICL architectures, which are aimed at realizing stable and high-performance hybrid multijunction cells. Preliminary implementations utilizing ALD $SnO_x$-based, CRL-free ICLs in P-O TSCs achieved only 22.31% efficiency[29,30], constrained by substantial electrical losses and suboptimal band alignment within the ICL. Crucially, the properties of ALD-grown $SnO_x$ can be finely tuned by modulation of deposition parameters (e.g., reaction time, temperature, and precursor selection[31-33]), offering a viable strategy to establish efficient electrical contacts while eliminating CRLs, thereby enhancing performance of CRL-free P-O tandem architectures.

Besides, a critical, currently underrepresented research aspect is the operational stability of P-O TSCs compared to the involved single-junction cells. While some studies have vaguely demonstrated that the architectural design of P-O tandem devices contributes to enhanced photostability, the mechanisms remain insufficiently addressed[14,15,18,19]. UV-visible light wavelength-dependent degradation is one of the leading degradation issues for OSCs[2,34], and recent work has shown that this fundamental degradation can be delayed or even prevented by using filters or light converters in the UV-visible regime[35,36]. Preliminary stability data of P-O TSCs indicate that the bottom WBG inorganic halide perovskite may play a similar protective role for the organic semiconductor materials by blocking harmful UV-visible light, thus contributing to improved operational stability[2,18,34,35]. However, there is a lack of in-depth experimental investigation into how the single-junction sub-cells of multi-junction cells influence each other's stability. Addressing this critical gap in understanding the long-term performance of P-O tandem devices is also a major focused aspect of this paper.

In this work, we present highly efficient and stable P-O TSCs employing a $C_{60}$/optimized ALD $SnO_x$/PEDOT: PSS ICL with significantly enhanced recombination efficiency. The key component of the optimized ICL, which eliminates the need of an additional CRL to achieve high transparency and reduced electrical losses, is a modified ALD $SnO_x$ layer that is deposited by sequentially using $H_2O$ and $H_2O_2$ with high reactivity as oxygen sources. A thin, $H_2O$- processed ALD $SnO_x$ film is deposited prior to the $H_2O_2$-converted $SnO_x$ film to protect the underlying films from the damage caused by $H_2O_2$ vapor. The resulting $H_2O$-/$H_2O_2$-processed ALD-$SnO_x$ film exhibits greatly improved electrical properties and forms an excellent quasi-ohmic contact with PEDOT: PSS in comparison to the conventional pure $H_2O$-processed $SnO_x$/PEDOT: PSS combination. Accordingly, the optimal ICL displays superb protection capability, effective charge recombination efficiency, and high NIR transmittance, which enables P-O TSCs to achieve an average efficiency of 25.12%, along with a *V*oc of 2.11 V and an outstanding short-current density (*J*sc) of 15.38 mA $cm^{-2}$. Additionally, we observe that the commonly used $C_{60}$ film undergoes recrystallization over operation time, leading to the formation of structures with grain boundaries, which serve as bi-directional migration channels for halide ions from the adjacent perovskite layer and the top metal Ag electrode. More interestingly, a compelling experimental study demonstrates that integrating the polymer organic layers within the tandem configuration largely suppresses the interdiffusion of mobile ions and prevents electrode corrosion. This integration establishes a mutual stabilization mechanism between the two single-junction sub-cells of the tandem configuration: the capping polymer organic sub-cell effectively suppresses ion-induced degradation, while the halide perovskite layer protects the organic absorber layer from the spectrally induced degradation. As a result, our target P-O TSCs show distinguished operational photostability, retaining 91.68% of initial efficiency after 1000 hours of continuous exposure to metal halide lamp (MHL) without any UV filter, which is significantly more stable than these single-junction counterparts. Moreover, these P-O TSCs exhibit extraordinary anti-fatigue behavior after a prolonged diurnal (12/12-hour) cycle test for 2067 hours.

**Characterization of ALD $SnO_x$ Films with Different Reactants and Device Performance**

**Figure 1a** illustrates the schematics of the ALD process (details are described in the Supplementary part). The experimental details of the ALD process based on $H_2O$ and $H_2O_2$ are schematically depicted in **Scheme S1**. Different oxygen reactants have varying reactivities, which determine the electronic properties of the $SnO_x$ layer[33,37]. Initially, the electrical properties of ALD-processed $SnO_x$ films with pure $H_2O$, $H_2O_2$, and $H_2O$/$H_2O_2$ reactants were examined in a lateral architecture. As depicted in **Figure S1b**, current-voltage characteristics of the Ag/$SnO_x$/Ag devices display a linear response for all $SnO_x$ films. The $H_2O$-derived $SnO_x$ film exhibits the lowest electrical conductivity, while the pure $H_2O_2$-based $SnO_x$ film shows the highest electrical conductivity. The electrical conductivities of two $H_2O$/$H_2O_2$-converted $SnO_x$ films with different combined cycles fall between them. Additionally, **Figure S1** shows that the $H_2O_2$-derived $SnO_x$ film has a smaller optical absorption tail than pure $H_2O$-derived $SnO_x$ film, indicating a reduction in the density of defect states present in the energy bands of the $H_2O_2$-derived $SnO_x$ film, which contributes to the improved electrical conductivity of bulk films[38]. Additionally, the presence of these gap states is confirmed at the $SnO_2$ surface/interface and further revealed to be Sn 5s lone pairs by the Al $K_\alpha$ XPS valence band spectra, as shown in **Figure S2a**[39,40]. Band energy diagrams of two types of $SnO_x$ films were constructed by combining UV-vis absorption spectroscopy with ultraviolet photoelectron spectroscopy (UPS) (**Figures S2b-2c**). The results suggest that the Fermi level moves closer to the conduction band minimum (CBM) for the $H_2O_2$-derived $SnO_x$ film, indicating the enhanced n-type doping behavior (**Figure S2d**). Next, the efficacy of varying reactants-derived ALD $SnO_x$ films as ETLs was explored based on single-junction halide perovskite solar cells (SPVKs) with a structure of ITO/$NiO_x$/Me-4PACz/$Al_2O_3$/GABr/1.81 eV $Cs_{0.3}FA_{0.7}Pb(I_{0.6}Br_{0.4})_3$/GABr: F-PEAI/$C_{60}$/$SnO_x$/Ag. **Figure 1c** depicts the current-voltage (*J-V*) characteristics of these devices, while **Table S1** summarizes their corresponding device parameters. Compared to the control devices with pure $H_2O$-converted $SnO_x$, which have an average PCE of 15.11%, both devices combined with two different $H_2O$/$H_2O_2$-converted $SnO_x$ films deliver improved device performance, along with the enhanced *J*sc and fill factor (FF), ascribing to the improved electrical properties of these films. Particularly, the optimal SPVKs with $H_2O$ (70 cycles)/$H_2O_2$ (70 cycles) $SnO_x$ film achieve a hero efficiency of 16.67% with a *J*sc of 16.79 mA $cm^{-2}$, a *V*oc of 1.26 V, along with an *FF* of 78.66%. It is noteworthy that the $H_2O$/$H_2O_2$-converted $SnO_x$ film with a larger number of $H_2O_2$ cycles (110 cycles) shows a tendency towards a slightly reduced performance of 15.99%. We further underline the relevance of processing a thin $H_2O$-derived $SnO_x$ layer beneath the $H_2O_2$ one. Despite the better n-type character of pure $H_2O_2$-derived $SnO_x$ films, corresponding SPVKs exhibit a significantly reduced performance of 12.42%, caused by a massive FF loss (61.88%) and a remarkable *V*oc loss (1.22 V), likely due to the degradation effect of $C_{60}$ and/or perovskite film from the diffusion of $H_2O_2$ and/or direct exposure to $H_2O_2$ atmosphere. These results demonstrate that a properly thick $H_2O$-converted $SnO_x$ film is beneficial in protecting the underlying sensitive layers prior to depositing a $H_2O_2$-derived $SnO_x$ film. The integrated *J*sc values from external quantum efficiency (EQE) measurements are consistent with those from *J-V* scans performed under the AM 1.5G solar simulator, depicted in **Figure 1d**. The statistical distributions of these device performances are illustrated in **Figure 1e**, demonstrating excellent reproducibility. Accordingly, the optimized $H_2O$/$H_2O_2$ (70/70 cycles)-converted ALD $SnO_x$ film was chosen for the following experiments.

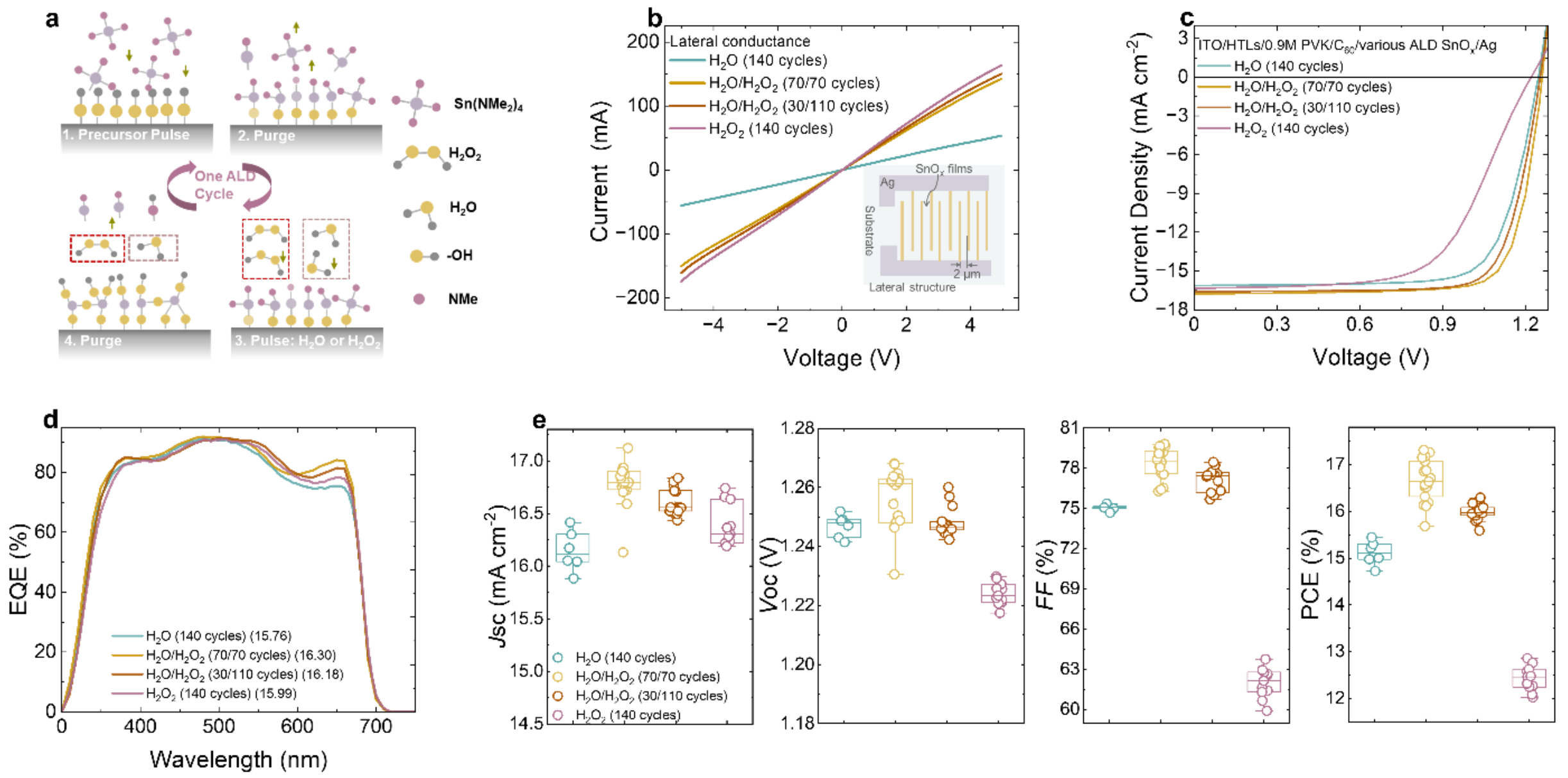

**Figure 1**. **Photovoltaic characteristics of SPVKs based on ALD $SnO_x$ processed from $H_2O$, $H_2O/H_2O_2$, and $H_2O_2$ reactants, respectively. a**, Schematic diagram of the ALD process, including precursor, reactant source pulse, and inert gas purge[41]. **b**, The lateral conductance measurements in a structure consisting of interdigitated Ag/$SnO_x$ films/interdigitated Ag. The inserted picture shows schematic diagram of the interdigitated finger electrode. **c, d,** and **e**, *J-V* curves measured with a reverse scanning direction, EQE spectra, and statistical performance (*J*sc, *V*oc, *FF*, PCE) of SPVKs based on various ALD $SnO_x$ films, respectively.

## Photovoltaic Performance of P-O TSCs

To explore the efficacy of the optimized ICL based on $H_2O/H_2O_2$-derived $SnO_x$ in a tandem configuration, a device stack of ITO/$NiO_x$/Me-4PACz/$Al_2O_3$/GABr/1.81 eV $Cs_{0.3}FA_{0.7}Pb(I_{0.6}Br_{0.4})_3$/GABr: F-PEAI/$C_{60}$/$SnO_x$/PEDOT: PSS/OSC/PDINN/Ag was built, as shown in **Figure 2a**. A cross-sectional scanning electron microscopy (SEM) image of the fresh tandem cells shows the intact multiple-layer structure, which suggests that the $SnO_x$ layer reliably protects the underlying WBG perovskite and $C_{60}$ layers from damage by the subsequent spin-coating layers, including the processing of a water-based PEDOT: PSS layer. For comparison, the $H_2O$-derived $SnO_x$-based ICL is used for the fabrication of the control P-O TSCs. Electrical and optical properties of ICLs are critical parameters to determine the performance of TSCs. **Figure S3** and **Figure 2b** demonstrate that the optimal ICL, the combined $H_2O/H_2O_2$-derived $SnO_x$ film, shows slightly higher transmittance in the UV range and significantly improved vertical conductivity in comparison to the control ICL. Additionally, compared to the control pure $H_2O$-converted $SnO_x$/PEDOT: PSS-based device which displays a curved *J-V* curve, the $H_2O/H_2O_2$-processed $SnO_x$/PEDOT: PSS sample exhibits improved quasi-ohmic contact, suggesting the formation of a more favourable interface contact and enhanced electrical transport properties. Next, the efficacies of these two ICLs were evaluated in the P-O TSCs by combining a highly efficient ternary organic absorber, PM6: L8BO: BTP-eC9, as the rear OSC sub-cell. The detailed chemical structures of the used organic semiconductors, the *J-V* curves, and EQE spectra of p-i-n single-junction ternary OSCs combining PEDOT: PSS as the hole transporting layer and PDINN as the ETL are depicted in **Figure S4.** The corresponding device performance parameters are summarized in **Table S2**, showing an average PCE of 17.47%. **Figure 2c** presents the *J-V* curves of the perovskite-ternary (P-T) TSCs with $H_2O$-derived $SnO_x$ and $H_2O$-/$H_2O_2$-derived $SnO_x$ based ICLs, respectively. The corresponding photovoltaics parameters are summarized in **Table S3**. By combining with the PM6: L8BO: BTP-eC9 OSC ternary rear sub cell, the control tandem device yields a PCE of 21.75%, with a $J$sc of 13.96 mA cm$^{-2}$, a $V$oc of 2.07 V, and an *FF* of 75.26%. In contrast, the target tandem device employing $H_2O$/$H_2O_2$-processed $SnO_x$-based ICL delivers an improved PCE of 23.53%, along with a $J$sc of 14.58 mA cm$^{-2}$, a $V$oc of 2.08 V, and an *FF* of 77.58% under the reverse scan, and their corresponding *JV* curves exhibit negligible hysteresis effect. The target P-T TSC shows the main improvement in $J$sc and *FF*. Next, a small amount of [70]PCBM was added to the ternary photo absorber to optimize the morphology of the rear OSC layer, enabling balanced charge transport and reduced nonradiative recombination, which lead to overall improved device parameters (**Figure S4** and **Table S2**)[42]. Accordingly, the tandem involving the quaternary OSC achieves an impressive average PCE of 25.12%, with a notably improved $J$sc of 15.38 mA cm$^{-2}$, a full $V$oc of 2.11 V, and an *FF* of 77.40% (**Figure 2d**). Notably, the significant enhancement in $J$sc value of the perovskite-quaternary OSC TSC (P-Q TSC) is attributed to the combined effects of morphological optimization of the quaternary rear OSC and reduced optical loss from the use of an anti-reflection film during the *JV* measurement. The perovskite-quaternary OSC-based TSC (P-Q TSC) displays excellent current matching between the front and rear sub-cells, as estimated from the integrated EQE spectrum (**Figure 2e**). Moreover, the optimal P-Q TSC exhibits a stabilized steady power output of 25.11% (**Figure 2f**). The performance statistics of P-Q TSCs verify the advantage of the ICL with $H_2O$-derived/$H_2O_2$-processed $SnO_x$ for highly efficient P-Q TSCs (**Figure 2g** and **Figure S5**). Additionally, **Figure S6** highlights the unipolar electron transport property of $H_2O$-/$H_2O_2$-derived $SnO_x$ as evidenced by a $C_{60}$/$H_2O$-derived and $H_2O_2$-derived $SnO_x$ ICL, yielding a tandem performance of only 11.41%. **Figure 2h** illustrates the $J$sc evolution *v*ersus $V$oc of state-of-the-art P-Q TSCs incorporating different kinds of ICLs with or without extra CRLs. This comparison highlights the significance of the optimal ICL architecture in minimizing optical and electrical losses.

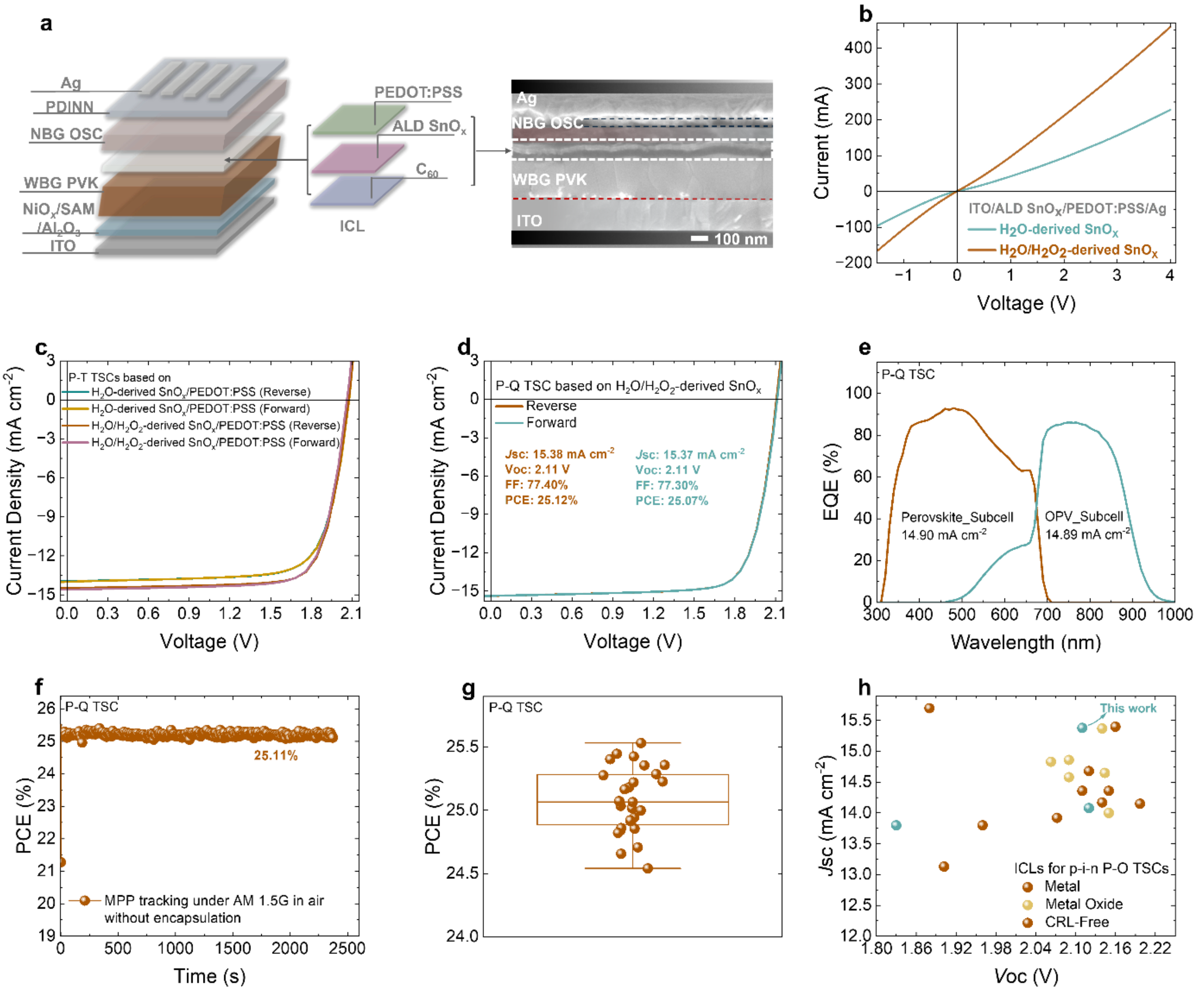

**Figure 2. Device architecture and performance of PO-TSCs. a**, Schematic diagram showing the p-i-n perovskite/organic TSCs and the corresponding cross-sectional SEM diagram. **b,** *J-V* characteristics of the diode devices with the structure of ITO/$H_2O$-derived and $H_2O$-derived/$H_2O_2$-derived $SnO_x$/PEDOT: PSS/Ag. **c**, *J-V* characteristics of P-Q TSCs with $H_2O$-derived $SnO_x$ and $H_2O$-derived/$H_2O_2$-derived $SnO_x$-based ICLs, respectively. **d** and **e,** *J-V* characteristics and the EQE spectra of P-Q TSC based on $H_2O$-derived/$H_2O_2$-derived $SnO_x$ based ICLs with reverse and forward scanning directions. Self-adhesive anti-reflection films were used in the measurement of P-Q TSCs. **f**, Stabilized PCE of the P-Q TSCs with $H_2O$-derived/$H_2O_2$-derived $SnO_x$ ICL. **g**, Statistical data of 28 P-Q TSCs measured with a reverse *J-V* scanning direction. **h**, *J*sc evolution versus *V*oc of P-Q TSCs with various ICLs with or without extra CRLs. A more detailed summary can be found in **Table S4**.

## Operational Device Stability of P-O TSCs

Both WBG halide perovskites and organic NFA-based OSCs are known to suffer from their respective leading degradation mechanisms. OSCs are prone to so-called burn-in effects, which are either correlated with a microstructure instability or spectrally sensitive degradation mechanism. WBG perovskite devices are susceptible to reactions between metal electrodes and mobile halide species, which can induce electrode corrosion and broad tail states in the perovskite layer, respectively[43-45]. The configuration of the P-O tandem device structure is speculated to offer a unique opportunity to mitigate or even to resolve these leading degradation mechanisms. On the one hand, capping the perovskite layer with the finely tuned organic semiconductor absorber layer, in combination with a well-designed of the ICL, can block bi-directional diffusion of halide ions and electrode ions and thus prevent corrosion of the top metal electrode, which is Ag in our case. On the other hand, the optical filter behavior of the perovskite front cell, absorbing all high-energy radiation below 600 nm, is expected to stabilize the OSCs from spectrally sensitive degradation[25,26]. Thus, we designed a comparative experimental campaign, allowing us to side-by-side follow operational stability evolution for the respective single-junction half cells versus their tandem configuration. First, long-term light operational stability of the single-junction quaternary PM6: L8BO: BTP-eC9: [70]PCBM OSCs (SQOSCs) and the WBG SPVKs, as well as the corresponding P-Q TSCs combined with the optimal $H_2O/H_2O_2$-derived $SnO_x$-based ICL was evaluated, where unencapsulated devices were subjected to continuous MHL illumination with stabilized powder-output tracking[46]. These unsealed devices without an extra UV filter were subjected to continuous illumination from MHL with power-output tracking. **Figure 3a** and **Figure S7** illustrate that both SPVKs and SQOSCs only preserve 52.83% and 80.49%, respectively, of their respective initial efficiency under short-circuit mode, while the target tandem devices maintain 91.68% of their original performance after 1000 hours under short-circuit mode (**Figure 3a**). This result strongly highlights the remarkable opportunity of the P-O tandem configuration in boosting device stability beyond both investigated single-junction counterparts. It is worth pointing out the tandem's temporal behavior under continuous operation is primarily governed by that of the perovskite sub-cell. Additionally, the lifetime reproducibility of the tandem devices was studied in different experimental runs and overall was found to be excellent (**Figures. S8-S9**). Additionally, the stability of the tandem device was investigated under open-circuit conditions with continuous illumination from the MHL illumination, retaining nearly 90% of its initial efficiency after aging near 1000 hours (**Figure S10**). Besides, it is noted that the SPVKs with an additional annealing-treated PEDOT: PSS polymer film on top demonstrate to a slower degradation rate compared to the control SPVKs (**Figure S11**)[47]. This suggests that while the additional thin PEDOT: PSS polymer film can delay ion diffusion, it cannot fully prevent it due to its limited thickness, implying the crucial role of organic semiconductor layer in suppressing the ion diffusion in the tandem configuration.

We next turn our attention to whether fatigue behavior of perovskite-based devices can be mitigated by the P-O tandem configuration, which can be most relevantly investigated under day/night cyclic operation. Under these conditions, perovskites often exhibit degradation during the day, recovery during the night (Type I) and/or degradation during the night, recovery during the day (Type II), which depends on the status of device degradation during a cycle[46,48,49]. To maximize the simulation of the diurnal condition, we further conducted 12-hour day/12-hour night cycling aging test on SPVKs, SOSCs, and P-O tandem devices following the "International Summit on Organic Photovoltaic Stability LC protocol" (ISOS-LC)[46]. **Figure S12** shows that SQOSCs exhibit no such fatigue behavior. As displayed **Figure 3b** and **Figure S13**, the WBG SPVKs display 33% performance loss, while the target tandem devices show marginal degradation after 61 aging cycles. The PCE evolution of SPVKs and P-Q TSCs in the representative cycles is further depicted in **Figure 3c**. In the case of SPVKs, which demonstrate evident fatigue behavior, the PCE initially decreases under illumination and recovers in the dark at the early cycles. The devices exhibit a noticeable "burn-in" effect, losing 22% of their PCE in the first cycle, mainly due to the obvious decrease in FF and $V_{OC}$ with a minor contribution of $J_{SC}$ loss within the first few hours (**Figure S14**)[50], while a full PCE restoration of the SPVKs is observed during the night, but followed by further degradation upon reillumination in the first 20 cycles, likely due to the light-induced ion migration[48]. Interestingly, the PCE dynamics of SPVKs shifts to Type II diurnal behavior in the later degradation stage (e.g. Cycle 50), indicating the emergence of new degradation mechanisms and the interplay among superimposed factors[49]. Both PCE restoration and degradation are observed in the aging operation, implying that reversible and irreversible degradation mechanisms co-exist in the WBG SPVKs[48]. Notably, both the PCE restoration rate and extent of SPVKs reduce gradually along with consecutive diurnal cycle, suggesting the increase of irreversible degradation mechanisms, such as irreversible migration of ionic species within the devices and corrosion at the metal electrode[43,44,49,51]. In contrast, the stability of tandem devices remains unchanged throughout the test, exhibiting extraordinarily enhanced anti-fatigue stability under cyclic light/dark stress conditions. This result underlines the successful stabilization of the perovskite half-cell in a tandem configuration with capped polymer organic layers on top. It is noted that the target tandem devices still operate at 96.80% of their initial performance, presenting literally burn-in free or light-soaking free operation even after a long light/dark cycle operation of 2067 hours (i.e., 86 cycles). This represents one of the best anti-fatigue stabilities for perovskite-based devices (**Figure S15**) so far. The overall stability results shown in **Figures 3a-3c** confirm the unique opportunity to stabilize perovskite cells by integrating them into a tandem configuration with a combination of the ICL and a polymer organic back sub-cell.

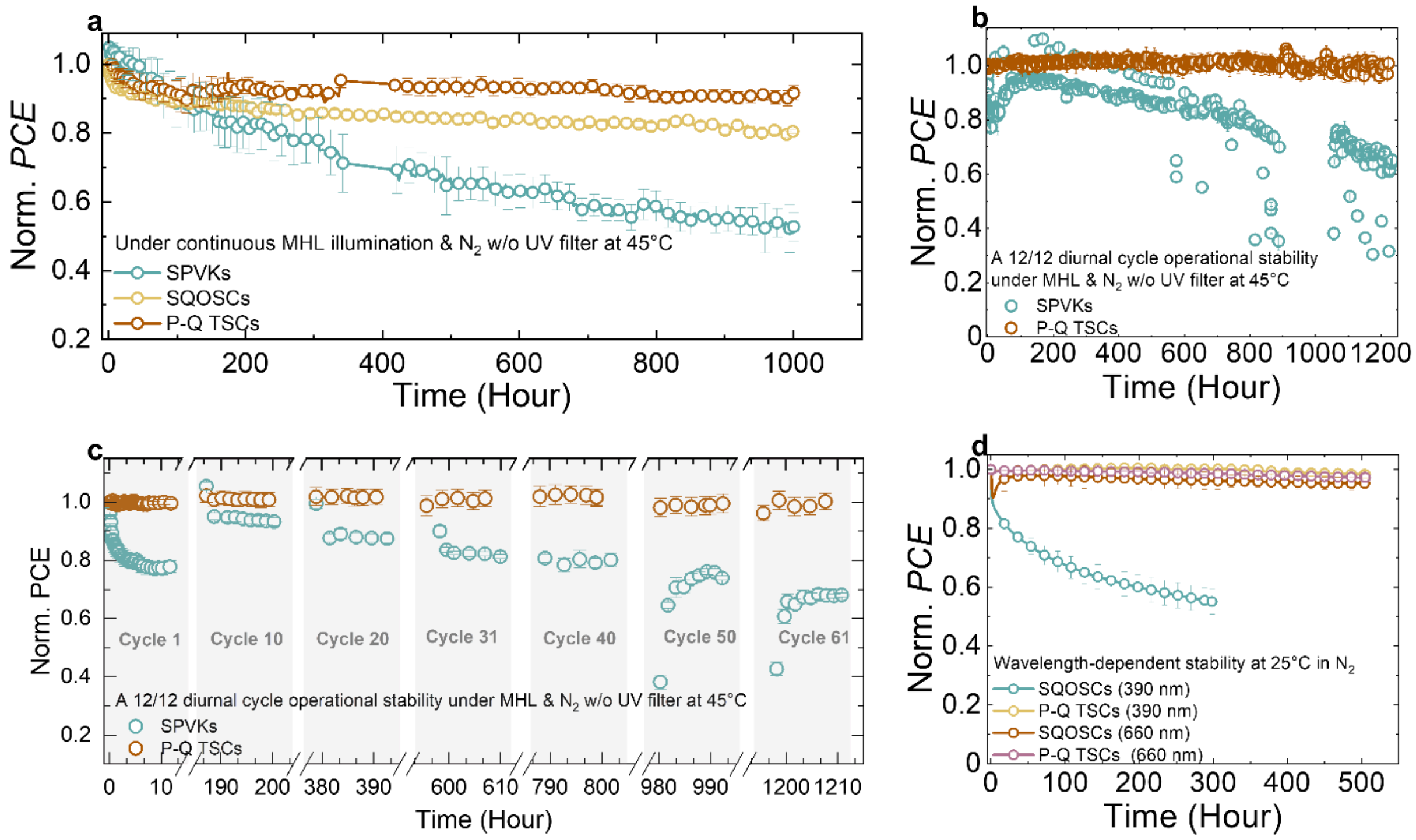

**Figure 3. Long-term device stability evaluation of P-Q TSCs. a,** Evolution of the normalized power output of unencapsulated SQOSCs, SPVKs, and P-Q TSCs under continuous MHL light irradiation and stabilized power-output tracking in a nitrogen flow. The stability of the P-Q TSCs was represented by the average data shown in **Figure S8**. **b**, **c** Simulating tracking over a 12/12-hour diurnal cycle operation for the SPVKs and P-Q TSCs. **d**, Evolution of normalized power output of SOSCs and P-Q TSCs illuminated under 390 nm (5 mW $cm^{-2}$) and 660 nm (100 mW $cm^{-2}$) monochromatic sources in a nitrogen flow.

## Preventing Spectrally Induced Degradation via the Perovskite Layer

Having established that the tandem configuration indeed shows superior operational stability as compared to the respective single-junction cells, we turn our attention to the underlying mechanism enhancing the stability of the organic sub-cell. In our tandem configuration, a 1.81 eV-PVK sub-cell with an absorption edge at approximate 710 nm is used. Our previous studies have shown that most organic solar cells demonstrate a dramatic increase in operational stability when protected from light below 600 nm[35]. In the tandem, the perovskite sub-cell is expected to efficiently filter out the damaging high-energy light for the rear organic sub-cell and allow it to operate stably in the benign absorption region above 600 nm[2,34]. Therefore, to evidence the stabilization mechanism of the organic sub-cell in a tandem configuration, we operated the respective SQOSCs and tandem-junction devices under two monochromatic LEDs peaked at 390 nm and 660 nm, respectively. **Figure 3d** and **Figure S16** demonstrate that both SQOSCs and P-Q TSCs maintain nearly unchanged performance at 660 nm irradiation with a 1-sun equivalent intensity, showing consistent behavior with the reported literature[34]. Nevertheless, when illuminated under a high-energy light at 390 nm (5 mW $cm^{-2}$), the SQOSCs degrade pronouncedly, while the corresponding tandem cells demonstrate a negligible performance loss, showcasing the excellent protective effect of the perovskite layer for the organic components.

**Blocking Bi-directional Migration Pathway via Dense Organic Layers**

Having demonstrated that the two sub-cells stabilize each other mutually, we turn to the question of how the polymer organic half-cell manages to stabilize the halide perovskite half-cell. It has been well-documented that the bi-directional migration of halides or halogen species and silver causes irreversible chemical corrosion to the electrodes and the depletion of halides in the perovskite layers, limiting the long-term operational stability of SPVKs[44,52,53]. Our P-Q tandem architecture exhibits extraordinary stability under illumination, indicating the effective suppression of bi-directional ion migration of the perovskite and silver electrode within the devices. To verify the effectiveness of the organic capping layer in suppressing the bi-directional ion migration of perovskite and Ag electrode during device operation, we prepared cross-sectional lamellae of SPVKs and target P-Q TSCs before and after ageing under continuous 1000 hours-operation and conducted scanning transmission electron microscopy (STEM) and energy dispersive X-ray spectroscopy (EDXS) to straightforwardly shed light on the superior long-term stability mechanism of P-Q TSCs over the single-junction perovskite device. **Figure 4a** and **Figure S17** depict that perovskite-related elements, such as Br, I, and Pb, are homogenously distributed within the perovskite layer of the fresh SPVK sample. This contrasts sharply with the elemental distribution observed in the SPVK sample after 1000 hours of light-induced degradation. As shown in **Figure 4b** and **Figure S18**, a massive accumulation of Br is revealed at the Ag electrode, accompanied by noticeable traces of I and even a small fraction of Pb. These findings clearly evidence that the thermally-evaporated $C_{60}$ film and the 20 nm-thick ALD $SnO_2$ layer fail to act as effective barriers for the diffusion of Br and I, which are eventually scavenged by Ag ions to form metastable or stable metal halide complexes[44]. Furthermore, we note that in the elemental signal profiles, the Br signal is much higher than that of I. The complementary amount of Br relative to I and Pb signals at the Ag layer indicates that the diffusivity of $Br^-$ ions across the $C_{60}$ layer toward the Ag electrode after decomposition is higher than that of the $I^-$ ions. Interestingly, we also observed a few distinct flocculent-like features, vertically distributed in certain regions in the $C_{60}$ layer, appearing as the bright flocculent-like features in the HAADF-STEM images of the aged SPVK device (**Figure 4b** and **Figure S19b**), suggesting the presence of heavier elements in these regions. To gain further insights into the flocculent-like features within the aged $C_{60}$ film, we examined the sum EDXS spectra from a region of flocculent-like feature (color coded and marked into area #1) and compare it with a region in its direct vicinity (color coded and marked into area #2), as shown in **Figure 4c.** The spectra from area #1 clearly reveals the existence of Pb and I with traces of Br and Ag. The appearance of Ag in the $C_{60}$ layer demonstrates the deep inward diffusion of the Ag into the perovskite layer during the ageing operation under illumination. In contrast, in area 2#, which lacks the visible flocculent-like feature, no detectable signals characteristic of either perovskite-related or Ag species can be observed. The corresponding EDXS maps (**Figure S20**) support the conclusion, confirming that the $C_{60}$ films form diffusion channels for ion migration during the solar cell operation. One hypothesis is that the $C_{60}$ film undergoes recrystallization during ageing operation, leading to the development of these structural features. To further verify this point, we took a close look into the $C_{60}$ layer using the phase contrast scanning transmission electron microscopy (STEM) bright-field (BF) imaging, which allowed us to best reveal the structure changes before and after ageing treatment. Our analysis reveals a clear difference in the crystallinity levels between the fresh and aged $C_{60}$ layers. For the $C_{60}$ film in the fresh SPVK device, the contrast of image in **Figure 4d**, along with the halo ring in Fast Fourier Transform (FFT) image demonstrate a basically amorphous phase. Ordered domains are occasionally observed with very weak lattice contrast. In contrast, the morphology of the $C_{60}$ layer in the aged SPVK sample is characterized by different crystalline domains (**Figure 4e**). The corresponding FFT image clearly shows sharp spots from lattices arranged into at least 5 discernable rings of different radii, suggesting higher degree of crystallinity, whereas the ring shape of the FFT image suggests the random orientation of crystalline $C_{60}$ film in imaged field of view in the aged film. The dark-trace, flocculent-like features in STEM-BF images (**Figure 4e**), representing the diffusion channels of the Pb, Br, I, and Ag, appear to follow grain boundaries of $C_{60}$ crystal domains, suggesting a grain boundary mediated diffusion mechanism. Having established the $C_{60}$ layer as an active element in regulating ion diffusion in the aged SPVK, we come back to the presumed suppression of ion diffusion in tandem cells. As expected, the tandem device demonstrates an entirely different behavior during the identical aging operation condition. **Figure 4g** and **Figure S22**, reveal that the characteristic mobile ions, Br, I, and Pb, remain largely “frozen” in the halide perovskite layer of the aged tandem device. Despite the slight ion redistribution within the bulk, the aged tandem device still maintains its perovskite structure and typical grain sizes, with the elemental distribution of mobile ions remaining similar to that of the fresh tandem device (**Figure 4f** and **Figure S21**). Additionally, while a few flocculent-like features are present in the aged tandem sample, they are significantly sparser compared to those in the SPVK sample (**Figures S19b, S19d**), suggesting the efficiently suppressed occurrence of ion migration. Furthermore, no detectable signals of mobile perovskite ions were observed in either the organic layer or the Ag top electrode. By further combining the device stability results of P-O tandem devices and SPVKs with the complete ICL (**Figure 3a** and **Figure S11**), it suggests that the capped polymer organic sub-cell serves as a primary permeation barrier, effectively preventing electrode corrosion caused by outward ion diffusion from the perovskite layer while also blocking the intrusion of Ag ions into the perovskite layer.

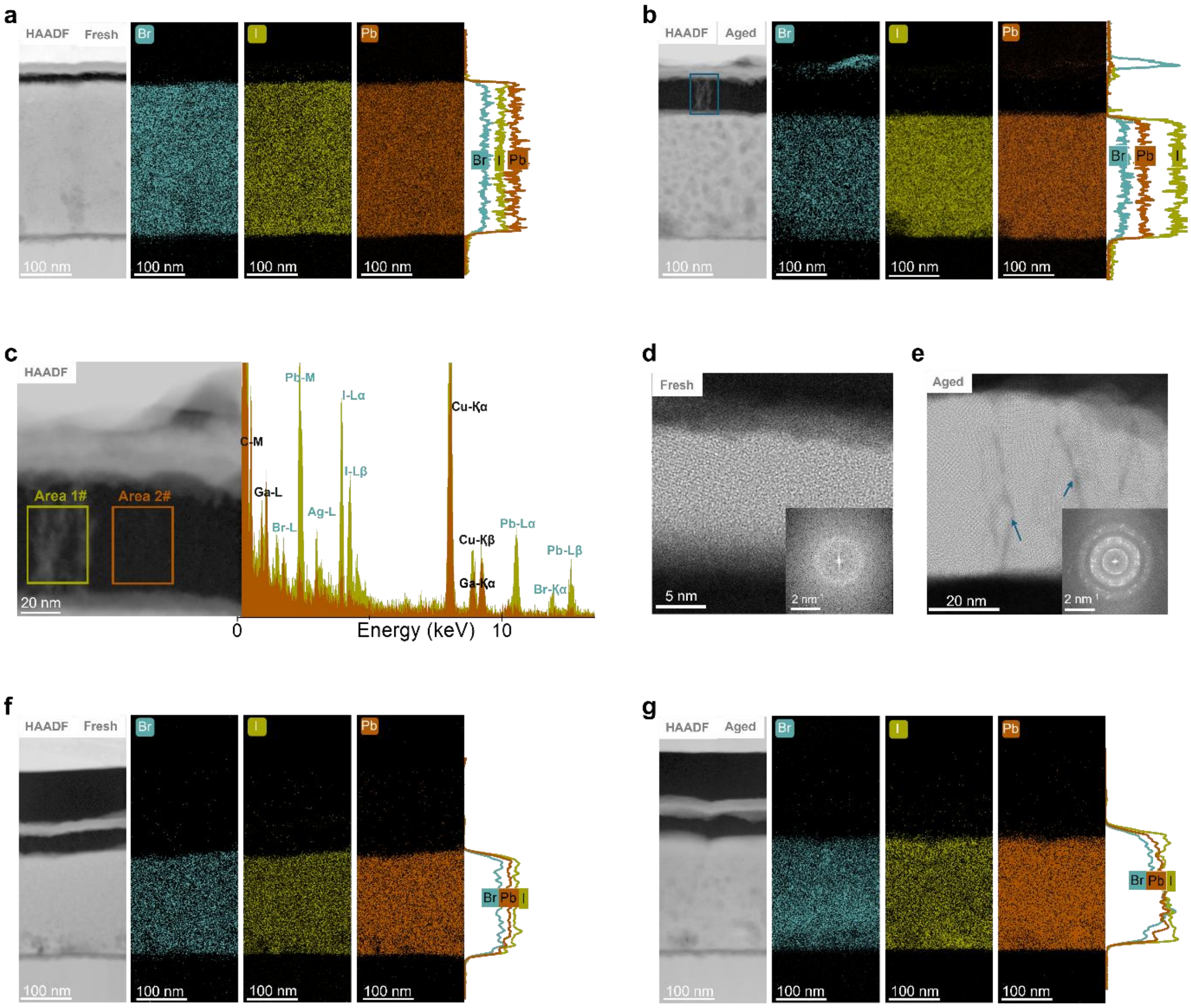

**Figure 4. Cross-sectional HAADF-STEM and EDXS characteristics. a, b,** Cross sectional STEM-HAADF images and the corresponding EDXS maps of Br, I, and Pb of fresh and aged SPVK, respectively. **c,** the zoom-in area of flocculent-like features in the aged $C_{60}$ film and the corresponding element signal spectra of two marked areas with and without flocculent-like features. **d, e,** STEM bright-field images of fresh and aged $C_{60}$ films for fresh and aged SPVK devices, respectively, with the FFT images as insets. **f, g,** Cross-sectional STEM-HAADF images and the corresponding EDXS maps of Br, I, and Pb of fresh and aged P-Q TSCs, respectively.

## CONCLUSIONS

In this work, we demonstrate outstanding performance and stability for hybrid P-O TSCs utilizing an advanced CRL-free ICL with a structure of $C_{60}$/ $H_2O$-derived and $H_2O_2$-derived ALD $SnO_x$/PEDOT: PSS. The ICL confers both excellent electrical, optical properties, and form a low-loss quasi-ohmic contact junction, which results in an averaged PCE of 25.12% (with a champion PCE of 25.53%) for P-O TSCs. Importantly, we provide critical insights into the enhanced stability of high-performance hybrid P-O TSCs. A notable finding is that the significant harmful bi-directional ion migration of the perovskite-related halide species and Ag ions occurs along the grain boundaries, which are formed in the $C_{60}$ film of single perovskite devices, resulting from recrystallization during ageing operation. Notably, the tandem configuration demonstrates unique stability advantages in efficiently regulating the ion interdiffusion by the integration of organic semiconductor layers, thus creating an inherently "self-stabilizing" system: the organic layers stabilize the perovskite sub-cell by suppressing ion diffusion-induced degradation, while the perovskite layer protects the organic sub-cell from spectral sensitivity-induced degradation. The simultaneous dual stabilization protection mechanism ensures exceptional long-term operational photostability for P-Q TSCs, reserving over 91% of their initial performance after 1000 hours of continuous light exposure under MHL at 45 °C; additionally, it enables negligible fatigue behavior after 86 operational 12/12-hour diurnal cycles (2067 hours), remarkably surpassing the stability of both single-junction cells. This work fully demonstrates the potential of P-O tandem configuration with inherent stability, alongside improved efficiency, offering a compelling prospect for intensifying future research into P-O tandem technologies. Additionally, uncovering the limitations of the commonly used $C_{60}$ film offers valuable insights for exploring more durable interface materials for future advancements in both perovskite-based single-junction and tandem architectures.

## CONFLICT OF INTEREST

There are no conflicts to declare.

## ACKNOWLEDGMENTS

C.L. gratefully acknowledges the financial support through the Helmholtz Association in the framework of the innovation platform "Solar TAP". C.L., K.Z, S. Q., Z.P., C.H.L., J.T. and J.Z. are grateful for the financial support from the China Scholarship Council (CSC). Z.P., C.H.L., J.T. gratefully acknowledge funding of the Erlangen Graduate School in Advanced Optical

Technologies (SAOT) by the Bavarian State Ministry for Science and Art. A. V. gratefully acknowledges the Slovak Research and Development Agency under the Contract no. APVV-23-0462. J.B. and J.E. acknowledge financial support from the Bavarian-Czech Academic Agency (BTHA), grant no. BTHA-JC-2024-2. P. W. gratefully acknowledges gratefully acknowledges funding through the "POPULAR" project (NO. 101135770). C.J.B. gratefully acknowledges the financial support through the "Aufbruch Bayern" initiative of the state of Bavaria (EnCN and SFF), the Bavarian Initiative "Solar Technologies go Hybrid" (SolTech), the DFG - SFB953 (project no. 182849149), and the DFG - INST 90/917-1 FUGG. MAA acknowledges financial support from the Fully Connected Virtual and Physical Perovskite Photovoltaic Lab (VIPERLAB) project. We thank the Energy Materials In-situ Laboratory Berlin (EMIL) for allowing us to perform lab-based UPS and XPS measurements.

**Author contributions**

C.L. and K.Z. conceived the idea of the work and designed the project. C.L., K.Z. and C.J.B. supervised the research. C.L. and K.Z. fabricated tandem devices. C.L. performed the optical and electrical characterizations, ALD processes, as well as device performance and stability measurements. X.Z., M.W., and E. S. performed FIB lamellae preparation, STEM, and EDXS measurements and greatly supported the analysis of these measurements. P.W. helped with the wavelength-dependent stability test. S.Q. measured SEM images. M. A., J.F., R.G.W., and M.B. performed and analyzed XPS and UPS measurements. J.E. and J.B. provided the ALD equipment support. Z.P. and Y.H. provided valuable assistance in stability data analysis by programing python script. T.H. provided the equipment support for the stability characterization. N.L., J.W., C.H.L., J.T., Z.P., and J.Z. provided valuable assistance in data analysis and curation. C.L. wrote the first draft of the manuscript. All the authors revised and approved the manuscript.

**Data availability**

All data which support this study are included in the published Article and its Supplementary Information.

**A Simultaneous Synergistic Protection Mechanism in Hybrid Perovskite-Organic Multi-junctions Enables Long-Term Stable and Efficient Tandem Solar Cells**

Chao Liu[1,2,#]*, Kaicheng Zhang[2,#]*, Xin Zhou[4], Mingjian Wu[4], Paul Weitz[2], Shudi Qiu[2], Andrej Vincze[5], Yuchen Bai[2], Michael A. Anderson[6,7], Johannes Frisch[6,7], Regan G. Wilks[6,7], Marcus Bär[1,6,7,12], Zijian Peng[2,9], Chaohui Li[2,9], Jingjing Tian[2,9], Jiyun Zhang[2], Jianchang Wu[1], Jonas Englhard[8], Thomas Heumüller[2], Jens Hauch[1], Yixing Huang[10], Ning Li[1,11], Julien Bachmann[8], Erdmann Spiecker[4], and Christoph J. Brabec[1,2,3]*

[1]Helmholtz-Institute Erlangen-Nürnberg for Renewable Energy (HI ERN), Immerwahrstraße 2, 91058 Erlangen, Germany

[2]Institute of Materials for Electronics and Energy Technology (i-MEET), Friedrich-Alexander-Universität Erlangen-Nürnberg (FAU), Martensstr. 7, 91058 Erlangen, Germany

[3]Institute of Energy Materials and Devices (IMD-3), Forschungszentrum Jülich GmbH, Wilhelm-Johnen-Straße 52428 Jülich, Germany

[4]Institute of Micro- and Nanostructure Research & Center for Nanoanalysis and Electron Microscopy (CENEM), Friedrich-Alexander-Universität Erlangen-Nürnberg (FAU), IZNF, Cauerstr. 3, 91058 Erlangen, Germany

[5]International Laser Centre SCSTI, 84104 Bratislava, Slovak Republic

[6]Department Interface Design, Helmholtz-Zentrum Berlin für Materialien und Energie GmbH (HZB), Albert-Einstein-Str. 15, 12489 Berlin, Germany

[7]Energy Materials In-situ Laboratory Berlin (EMIL), Helmholtz-Zentrum Berlin für Materialien und Energie GmbH (HZB), Albert-Einstein-Str. 15, 12489 Berlin, Germany

[8]Chemistry of Thin Film Materials, Department of Chemistry and Pharmacy, Friedrich-Alexander-Universität Erlangen-Nürnberg (FAU), IZNF, Cauerstr. 3, 91058 Erlangen, Germany

[9]Erlangen Graduate School in Advanced Optical Technologies (SAOT), Paul-Gordan-Straße 6, 91052 Erlangen, Germany

[10]Institute of Medical Technology, Health Science Center, Peking University, Xueyuan Rd. 38, Haidian District, 100083, Beijing, China

[11]Institute of Polymer Optoelectronic Materials & Devices, Guangdong Basic Research Center of Excellence for Energy & Information Polymer Materials, State Key Laboratory of Luminescent Materials & Devices, South China University of Technology, Guangzhou, 510640, P. R. China

[12]Department of Chemistry and Pharmacy, Friedrich-Alexander-Universität Erlangen-Nürnberg (FAU), Egerlandstr. 3, 91058 Erlangen, Germany

[#]These authors contributed equally: Chao Liu, Kaicheng Zhang

*Correspondence:

c.liu@fz-juelich.de (C.L.),

kaichengzhang@zoho.com (K.Z.),
christoph.brabec@fau.de (C.B.)

## EXPERIMENTAL PROCEDURES

### Resource Availability

*Lead Contact*

Further information and requests for resources and materials should be directed to and will be fulfilled by the Lead Contact, Chao Liu (c.liu@fz-juelich.de), Kaicheng Zhang (kaichengzhang@zoho.com)

### Methods

**Materials**: The indium tin oxide (ITO) glass substrates were bought from Advanced Election Technology Co., Ltd. PEDOT: PSS (VPAL 4083) solution was purchased from Heraeus Clevios. PM6, L8BO, BTP-eC9, and PDINN were from Solarmer. [70]PCBM was purchased from Solenne BV. $C_{60}$ was purchased from Nano-C. Methylammonium iodide (MAI, 99.99%), Formamidinium iodide (FAI, 99.99%), Formamidinium bromide (FABr, 99.99%), and 4-Fluoro-Phenethylammonium-iodide (F-PEAI, > 99.99%) were ordered from Greatcell Solar Ltd. Lead iodide ($PbI_2$, 99.99%). [4-(3,6-dimethyl-9-H-carbazole-9-yl)butyl]phosphonic acid (Me-4PACz, > 99.0%) was obtained from TCI Co., Ltd. Aluminium oxide ($Al_2O_3$) dispersion (20 wt% in isopropanol (IPA)), lead bromide ($PbBr_2$, 99.999%), cesium iodide (CsI, 99.999%), cesium bromide (CsBr, 99.999%), lead chloride ($PbCl_2$, 99.999%), and nickel(II) nitrate hexahydrate ($Ni(NO_3)_2 \cdot 6H_2O$, 99.999%) and ethanolamine (> 98%) were purchased from Sigma Aldrich. Guanidinium bromide (GABr, 98%) and methylammonium chloride (MACl, 98%) were obtained from Xi'an Yuri Solar Co. Ltd. $Sn(NMe_2)_4$ (99%) was purchased from abcr GmbH and $H_2O_2$ (35%, stabilized) was purchased from Carl Roth. All the chemicals were used as received without further purification.

**Perovskite solutions**: The wide bandgap perovskite 0.9 M $Cs_{0.3}FA_{0.7}Pb(I_{0.6}Br_{0.4})_3$ with additional 1 mol% $Pb(I_{0.6}Br_{0.4})_2$ and 1 mol% $MAPbCl_3$ was prepared by mixing CsI, CsBr, FAI, FABr, MACl, $PbI_2$, $PbBr_2$, and $PbCl_2$ in DMF / DMSO (4/1) solvents. The precursor was stirred at 55 °C for 30 min and 25 °C for 2 h, following by filtering with 0.25 μm PTFE filter before using.

**Atomic layer deposition (ALD) of $SnO_2$ film**: The $SnO_x$ layer was prepared by ALD in a commercial Gemstar-6 XT ALD reactor equipped with a Cobra BA 0100 C pump from Busch and with $N_2$ as carrier gas. $Sn(NMe_2)_4$ and $H_2O$ (or $H_2O_2$) were used as precursors and maintained in stainless steel bottles at 65 °C and at room temperature, respectively. The chamber of the reactor was heated up to 100 °C. For the deposition of $H_2O$-derived $SnO_x$, the tin precursor was pulsed for 0.4 s into the chamber, stayed in there for 7 s (exposure), and then subsequently pumped away by applying a continuous $N_2$ flow and vacuum to the chamber for 10 s (purge). In the second half-cycle, $H_2O$ was pulsed into the reactor with pulse, exposure and purge times of 0.1 s, 7 s, and 20 s, respectively. 140 ALD cycles yielded a deposition of ca 20 nm $SnO_x$ film thickness, as determined by characterizing parallelly ALD-grown films on native Si(100) wafers via. For the $H_2O/H_2O_2$-derived $SnO_x$ film, $H_2O$-based ALD process was run first, then followed by the $H_2O_2$-based ALD process. The parameters of ALD process remain constant. The ALD cycles of $H_2O$-derived $SnO_x$ and $H_2O_2$-derived $SnO_x$ for the $H_2O/H_2O_2$-$SnO_x$ film are varied according to the experiment requirements.

**Single-junction WBG perovskite solar cells**: The pre-patterned ITO-coated glass substrates (25×25 $mm^2$, sheet resistance: 15 Ω/sq) underwent sequential cleaning by ultrasonication in acetone and isopropanol for 10 min each. Following this, After these substrates were treated with $O_3$-plasma for 3 min, 60 μL $NiO_x$ solution (250 mg $Ni(NO_3)_2·6H_2O$ dissolved in 60 μL ethanolamine mixed with 10 mL ethanol) was deposited on these ITO substrates at 5000 rpm for 30 s; subsequentially, thermal annealing treatment at 100 °C for 5 min and then 300 °C for 30 min were followed to form a thin $NiO_x$ layer. Then, the substrates were transferred into glovebox. Me-4PACz SAM solution (0.5 mg/mL in ethyl alcohol) was applied on $NiO_x$ film, which was left for 5 s and coated at 4000 rpm for 30 s, and then annealed at 100 °C hot plate for 5 min. Subsequently, a diluted $Al_2O_3$ dispersion (1:100 in IPA) was spin-coated on the SAM layer at 4000 rpm for 25 s and then heated at 100 °C for 5 min. Next, the 80 μL GABr solution (1 mg/ml in IPA) was spin-coated atop at 5000 rpm/20 s, followed by heating at 100 °C for 10 min. After cooling down to room temperature, 60 μL of a wide bandgap perovskite solution (0.9 M) was dispensed at the center of substrate and spin-coated in two steps: at 1000 rpm for 5 s and then 6000 rpm for 40 s. During the final 30 s, a $N_2$ flowing gun (pressure: around 0.2 mbar) was then applied to dry the solvents in the film. Subsequently, the film was directly placed on the hotplate with an annealing temperature of 100 °C for 10 min, during which the color transitioned to dark brown. A mixed passivation solution (0.5 mg GABr + 1 mg F-PEAI in 1 ml IPA) was deposited on the perovskite with a speed of 5000 rpm/20 s and followed by annealing at 100 °C for 5 min. Next, a 25 nm thick $C_{60}$ layer was thermally evaporated on the halide perovskite film, followed by the deposition of a 20 nm $SnO_x$ layer by the ALD process. Finally, the devices were finished by thermally evaporating a 100 nm Ag electrode with an active area of 3.85 $mm^2$.

**NBG organic solar cells**: The pre-patterned ITO-coated glass substrates (25×25 $mm^2$ with sheet resistance of 15 Ohm/sq) were cleaned by ultrasonication in acetone and isopropanol for 10 min each. Subsequently, PEDOT: PSS solution was spin-coated on the top of plasma-treated ITO with a speed of 5000 rpm/40s and then annealed at 140 °C/15min in the air. Then, the substrates were transferred into glovebox immediately. PM6: L8BO: BTP-eC9 (1:0.65:0.65 by weight ratio) or PM6: L8BO: BTP-eC9: [70]PCBM (1:0.65:0.65:0.2 by weight ratio) with a polymer concentration of 7 mg/ml in chloroform was deposited on the top of PEDOT: PSS film with a spin speed of 4500 rpm for 30 s. Additional 0.3% DIO by v/v was added into the active solution. The active layers were annealed at 100 °C for 10 min in the glovebox. Then PDINN (1mg/ml in methanol) was coated with a speed of 3000 rpm/30s. All films were deposited with a dynamic spin-coating method. Finally, the devices were finished by thermally evaporating a 100 nm Ag electrode with an active area of 3.85 $mm^2$.

**Perovskite/organic tandem solar cell fabrication**

The fabrication of the front perovskite sub-cells was similar to that of the single-junction perovskite solar cells. After the deposition of an ALD $SnO_x$ layer on the perovskite sub-cells, a diluted PEDOT: PSS solution (1:1 with $H_2O$) was deposited on the top of ALD SnO*x* layer. Then the whole devices were thermally annealed at 120 °C for 10 min and subsequently transferred into the glovebox immediately. The fabrication of the rear organic sub-cells was the same as the single-junction organic solar cells. Finally, the devices were finished by thermally evaporating a 100 nm Ag electrode with an active area of 3.85 $mm^2$.

**Scanning electron microscopy (SEM) images**: The cross-section SEM images of the SPVK and P-Q TSC devices were investigated using a Schottky field-emission SEM (JEOL JSM-7610F) with the acceleration voltage of 3 kV~ 10 kV.

***J-V* measurement**: *J-V* characteristics were measured with a Keithley source measurement unit and a WAVELABS SINUS-70 solar simulator, providing illumination with an AM 1.5G spectrum and light intensity of 100 mW $cm^{-2}$. The light intensity was calibrated with a standard crystalline Si device from Newport. A mask with an aperture area of 3.85 $mm^2$ was used to get *J-V* measurements. For the single junction perovskite devices, the reverse scan of *J-V* curves was from 1.60 V to -0.2 V with a step of 0.05 V and a delay time of 20 ms, while the forward scan was from -0.2 V to 1.60 V with same step and delay time. For P-O tandem devices, the reverse scan of *J-V* curves was from 2.80 V to -0.2 V with a step of 0.06 V and a delay time of 20 ms, while the forward scan was from -0.2 V to 2.80 V with the same step and delay time. A self-adhesive anti-reflection film was only used for the *J-V* measurement of perovskite-quaternary OSC tandem devices. The lateral finger-interdigitated electrodes from Siemens were used for lateral conductivity measurements of $SnO_x$ films with a voltage range of -5 V to 5 V with a step of 0.09 V. The electrical properties of ICLs with a structure of ITO/$SnO_x$/PEDOT: PSS/Ag were measured with a voltage range of -5 V to 5 V with a step of 0.09 V. All devices were measured in the air atmosphere at room temperature and humidity around 35%.

**External quantum efficiency (EQE) spectra**: EQE of devices were measured using an EQE measurement system assembled by Enli Technology (Taiwan) with a lamp certified by a standard silicon cell and chopper frequency of 133 Hz. There are two additional laser sources with wavelengths of 450 nm (CW450-05). and 808 nm (LDM808/3LJ) to provide bias light for EQE measurements of narrow bandgap and wide bandgap sub cells, respectively.

**X-ray and Ultraviolet Photoemission Spectroscopy (XPS and UPS) measurements**: Samples for XPS and UPS were prepared on ITO substrates and shipped in sealed $N_2$ capsules to HZB/EMIL for characterization. Measurements took place in the dark in ultra-high vacuum with a base pressure of 1 x $10^{-9}$ mbar using a PHOIBOS 150 analyzer arranged at the "magic-angle" geometry (~54.7°) relative to the excitation sources. The analyzer pass energy was set to 200 eV for XPS survey spectra, 20 eV for XPS core level spectra, and 2 eV for UPS spectra. XPS spectra were obtained using monochromatic Al $K_\alpha$ (1486.71 eV) excitation from a SPECS XR 50 Al-Ag twin-anode x-ray source and a FOCUS 500 monochromator. UPS spectra were obtained using the He I (21.22 eV) and He II (40.81 eV) excitations from a Specs UVS 10/35 gas-discharge lamp. The He I excitation was used for the measurement of the secondary electron cut-off and the He II excitation was used for the valence band regions due to the contributions of He I satellites and low-BE He II signals near the He I Fermi level. A bias of -10 V was applied to the samples during UPS measurements, and the samples were grounded with the instrument to align the Fermi levels during XPS measurements. The binding energy scale was calibrated to the Fermi edge of clean Au foil at 0 eV for both UPS and XPS.

**Scanning transmission electron microscopy (STEM) and energy dispersive X-ray spectroscopy (EDXS)**: The Cross-sectional lamellae of SPVKs and P-Q TSCs for STEMS investigation, were prepared following the standard lift-out procedure within a dual-beam FIB-SEM Helios NanoLab 660 (Thermo Fischer Scientific). To avoid the possible degradation of the devices while exposed to air environment, they were all vacuum-sealed and unsealed just prior to the FIB processing. The final lamellae were finished with ion-beam showering at 2k eV beam energy to ensure the ion-beam-induced damages was kept at a minimum level. The as-prepared lamella was immediately sent into TEM, limiting overall exposure to the ambient conditions less than 5min. STEM imaging and EDXS acquisition were conducted using a Thermo Fischer Scientific (a double Cs-corrected Titan Themis), equipped with a Super-X EDXS system and operated at 300 kV. High angular annular dark field (HAADF) imaging was performed with a convergence half-angle of 15 mrad, a camera length of 91 mm, and a capturing angular range of 61-200 mrad. EDXS measurements were implemented with a sampling size (i.e., pixel size) of 1.53 nm/pixel and a dwell time of 50.0 μs, utilizing a probe current of 130-150 pA. The acquired images and EDXS results were evaluated using TFS Velox software (version 3.12).

**Long-term device stability measurement**: The long-term operational stability tests were conducted in a nitrogen-flowing chamber. These tests involved aging the devices under 0.85-sun illumination provided by metal-halide lamps, which emitted light in the wavelength range of 400 to 800 nm. During the test, no UV filters were used. Data collection was automated, and a cooling system maintained the average temperature of the system at approximately 45 °C. For the diurnal cycle tracking, the light source was cut-off manually. For wavelength-dependent stability, the unencapsulated devices were aged in nitrogen-flowing chamber, using two monochromatic LEDs with a central wavelength of 390 nm and 660 nm, respectively. The *J-V* characteristics of tested devices were then recorded hourly under 1-sun illumination using a white LED source with an intensity of 100 mW $cm^{-2}$ automatically. During this testing, the temperature of the system was consistently maintained at 25 °C. The devices used in all stability tests did not have anti-reflection coating films.

$$OH_x^* + \mathrm{Sn(NMe_2)_4} \longrightarrow O_x\,Sn(\mathrm{NMe_2})^*_{4-x} + xHNMe_2 \uparrow \qquad (1)$$

$$O_x\,Sn(\mathrm{NMe_2})^*_{4-x} + 2\,H_2O_2 \text{ or } 2\mathrm{H_2O} \longrightarrow O_2Sn(\mathrm{NMe_2})^*_x + (4-x)HNMe_2 \uparrow + \mathrm{O_2} \uparrow \qquad (2)$$

**Scheme S1**. **Chemical reactions of the ALD process to form $SnO_x$ using TDMASn and $H_2O_2$ or $H_2O$.** DMA is the dimethylamino ligand, HDMA is dimethylamine, and *x* is the number of DMA ligands released during the TDMASn exposures.[1] The asterisks represent the surface species.

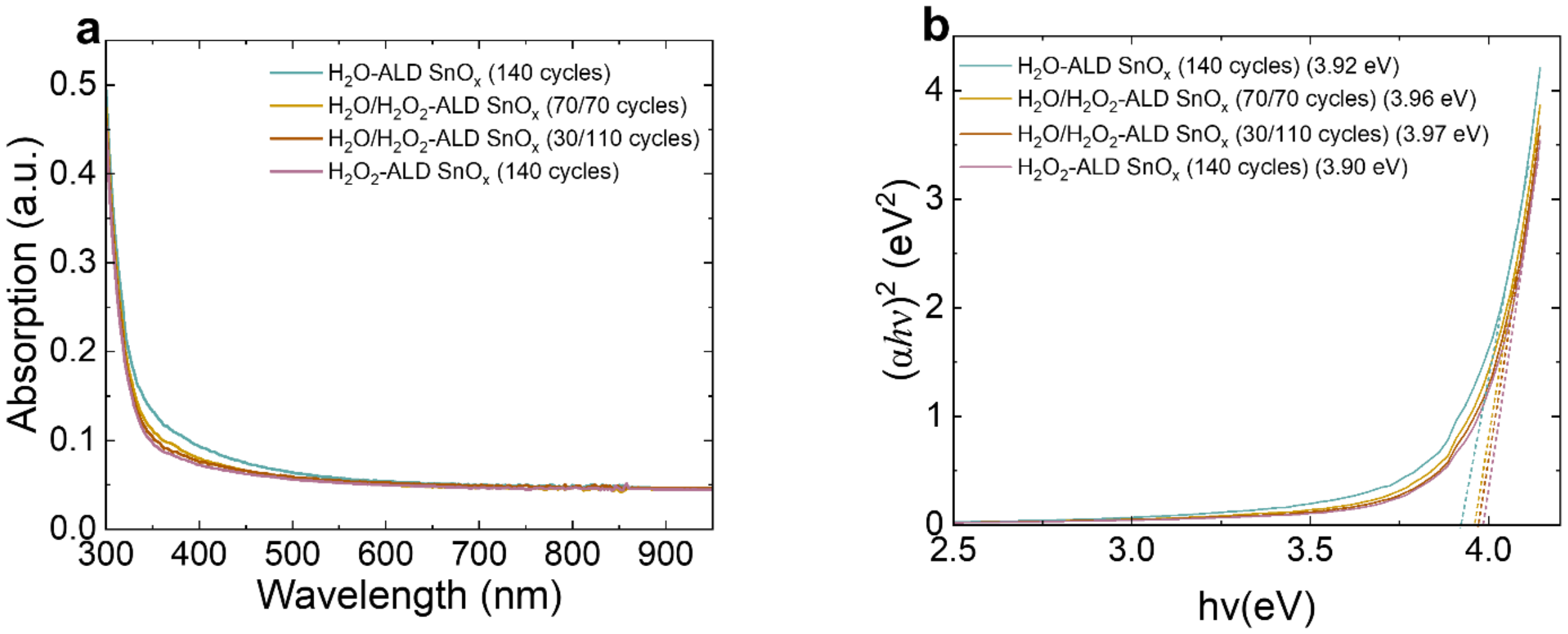

**Figure S1. UV-vis absorption and Tauc plots for various ALD $SnO_x$ films. a**, Absorption spectrum of pure $H_2O$-derived (140 cycles) and $H_2O_2$-derived $SnO_x$ (140 cycles) as well as two $H_2O/H_2O_2$-processed ALD-$SnO_x$ films with combined cycles of 70/70 and 30/110. **b**, The optical bandgaps of all involved ALD $SnO_x$ films, which were extracted using a linear fit to the absorption curve region.

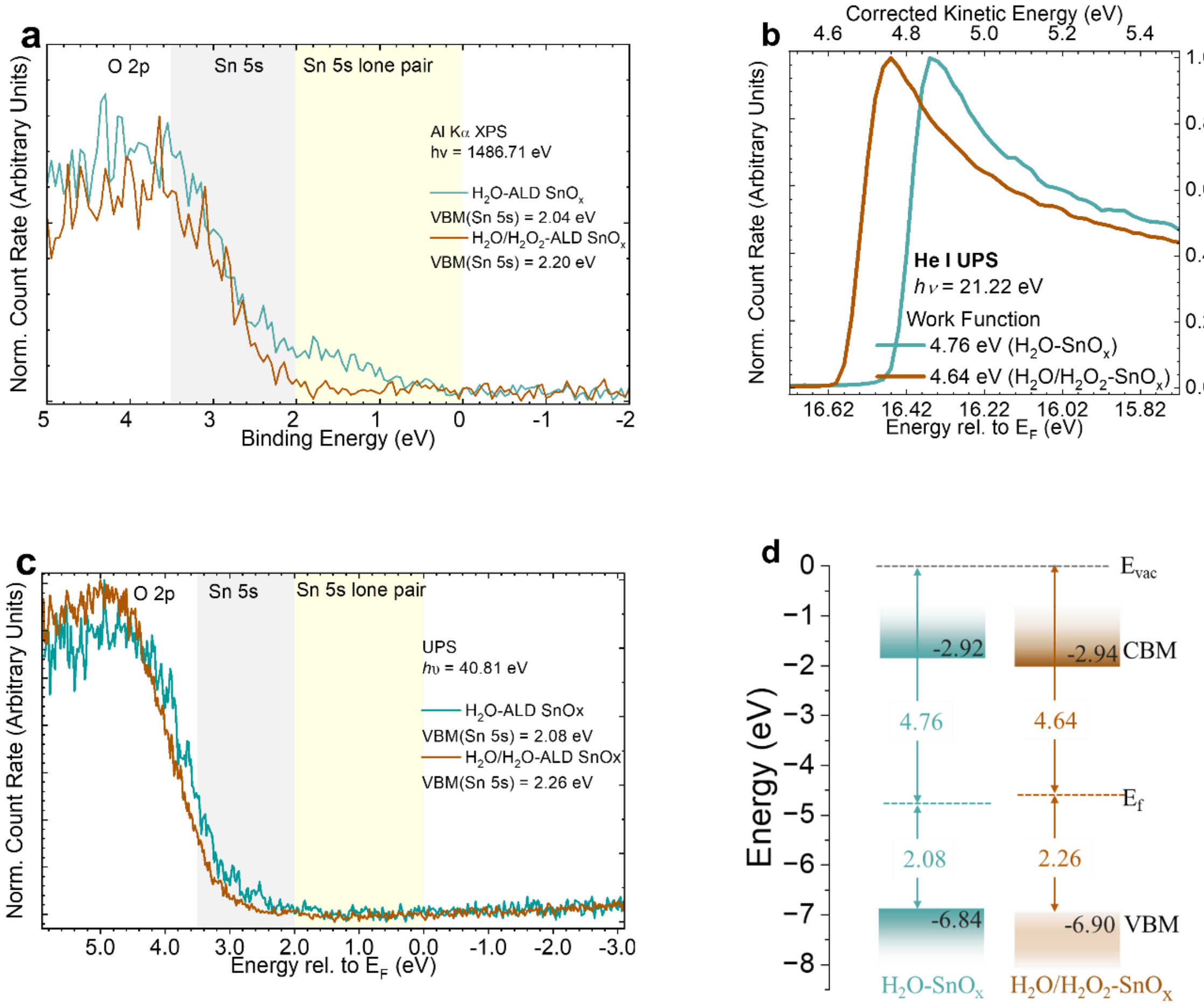

**Figure S2. XPS and UPS measurements for ALD $SnO_x$ films as well as corresponding energy alignment.** **a**, Al $K_\alpha$ excited valence band maximum (VBM) spectra of a $H_2O$-ALD $SnO_x$ and a $H_2O/H_2O_2$-ALD $SnO_x$ film. The positions of the VBM were derived by linear fits to low binding energy shoulder of the Sn 5s feature to be 2.04 ± 0.1 eV and 2.20 ± 0.1 eV, respectively. **b**, The secondary electron cut off for the $H_2O$-derived $SnO_x$ film and $H_2O/H_2O_2$-ALD $SnO_x$ film. The work functions of 4.76 ± 0.03 eV for $H_2O$-ALD $SnO_x$ and 4.64 ± 0.03 eV for $H_2O/H_2O_2$-ALD $SnO_x$ film were extracted using a linear fit to edge regions of the secondary electron cut off curve. **c**, He II excited valence band maximum (VBM) spectra and values of two ALD $SnOx$ films determined by linear fits to the Sn 5s feature feature, respectively. **d**, Energy level positions of the $H_2O$-$SnOx$ and the $H_2O/H_2O_2$-$SnO_x$ film based on a structure of ITO/$NiO_x$/Me-2PACz/$Al_2O_3$/GABr/perovskite/GABr: F-PEAI/$C_{60}$/$SnO_x$, where the VBM values were determined by the UPS He II measurements.

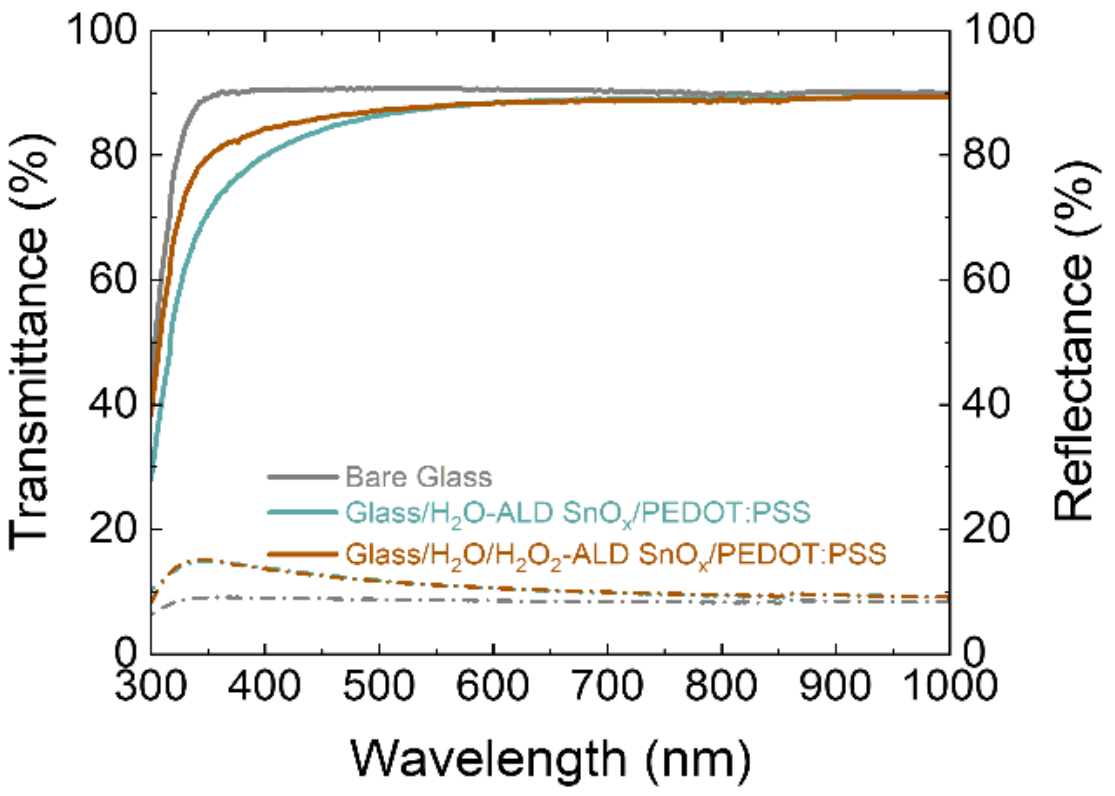

**Figure S3.** Transmittance and reflection spectra of $H_2O$-based $SnO_x$ and $H_2O/H_2O_2$-based $SnO_x$ ICLs.

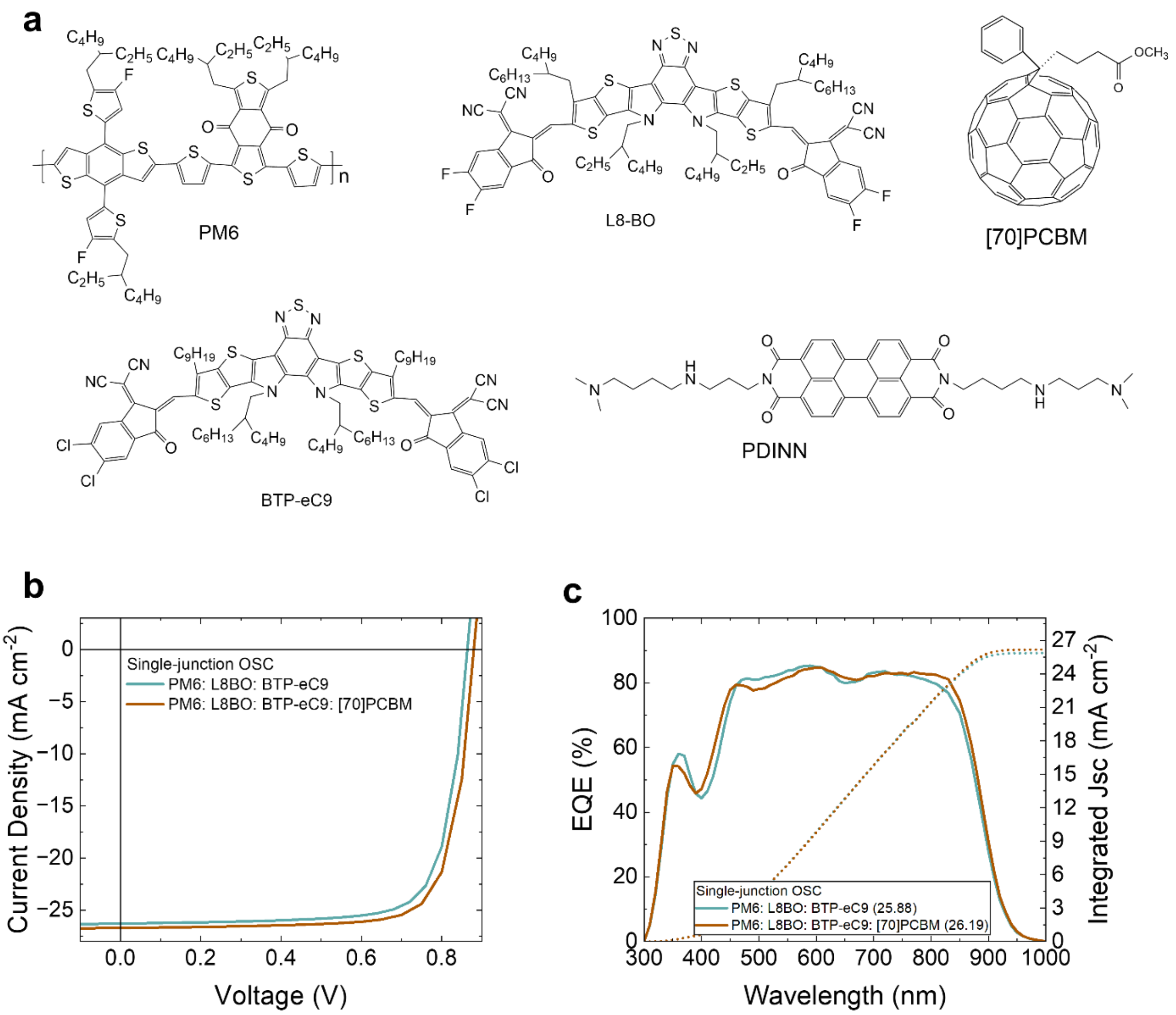

**Figure S4. Chemical structures, *J-V* characteristics and EQE spectra. a,** Chemical structures of the donor PM6, acceptors L8BO, BTP-eC9, and [70] PCBM as well as PDINN used in the ternary and quaternary organic photo-absorbers. **b, c,** *J-V* curves and corresponding EQE spectra of single-junction ternary and quaternary OSCs without anti-reflection film, respectively.

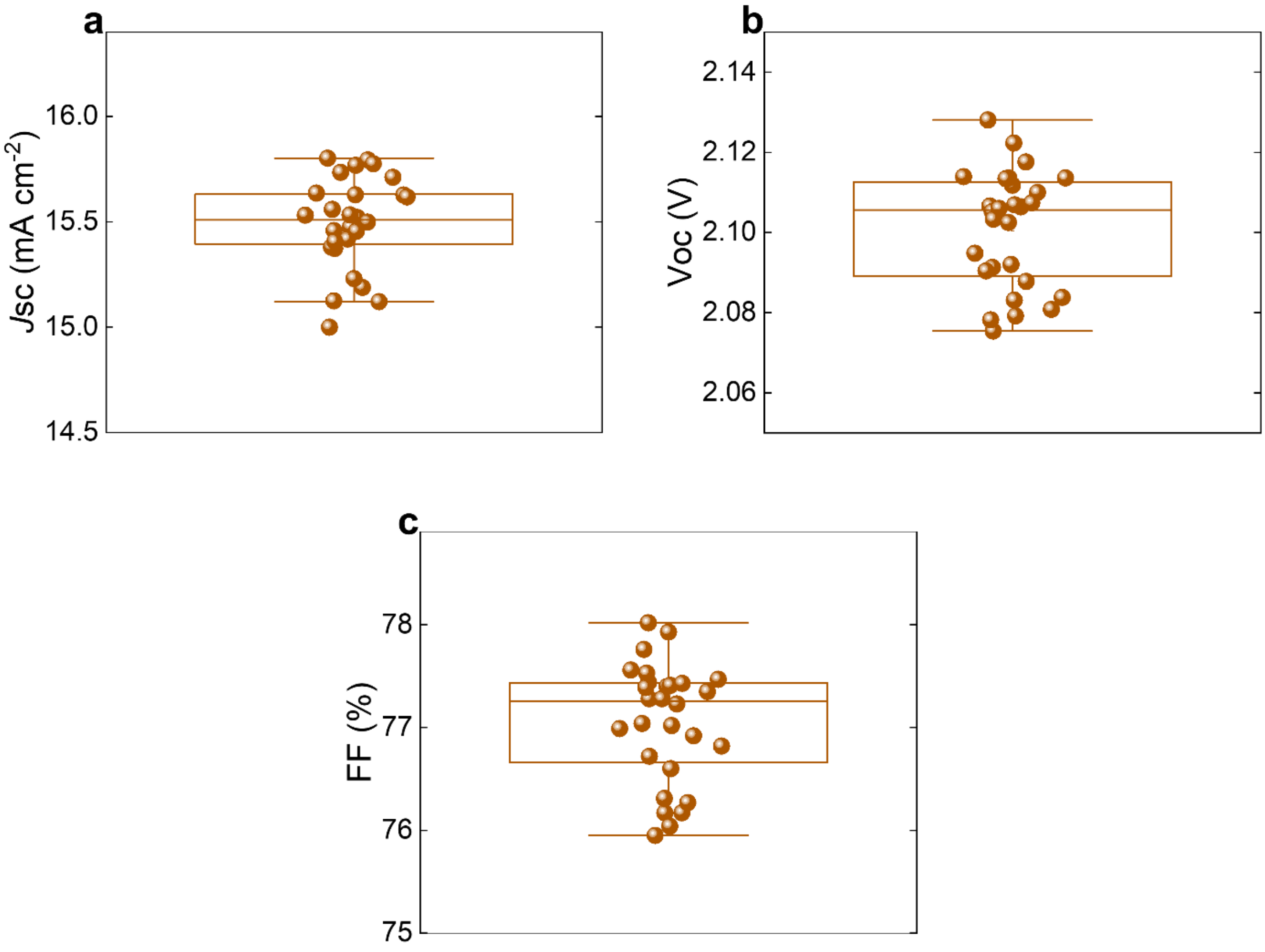

**Figure S5. Statistics of photovoltaic parameters for 28 P-Q TSCs with $H_2O/H_2O_2$-derived $SnO_x$ ICLs.** Statistics of **(a)**, *J*sc, **(b)**, *V*oc, and **(c)**, *FF*.

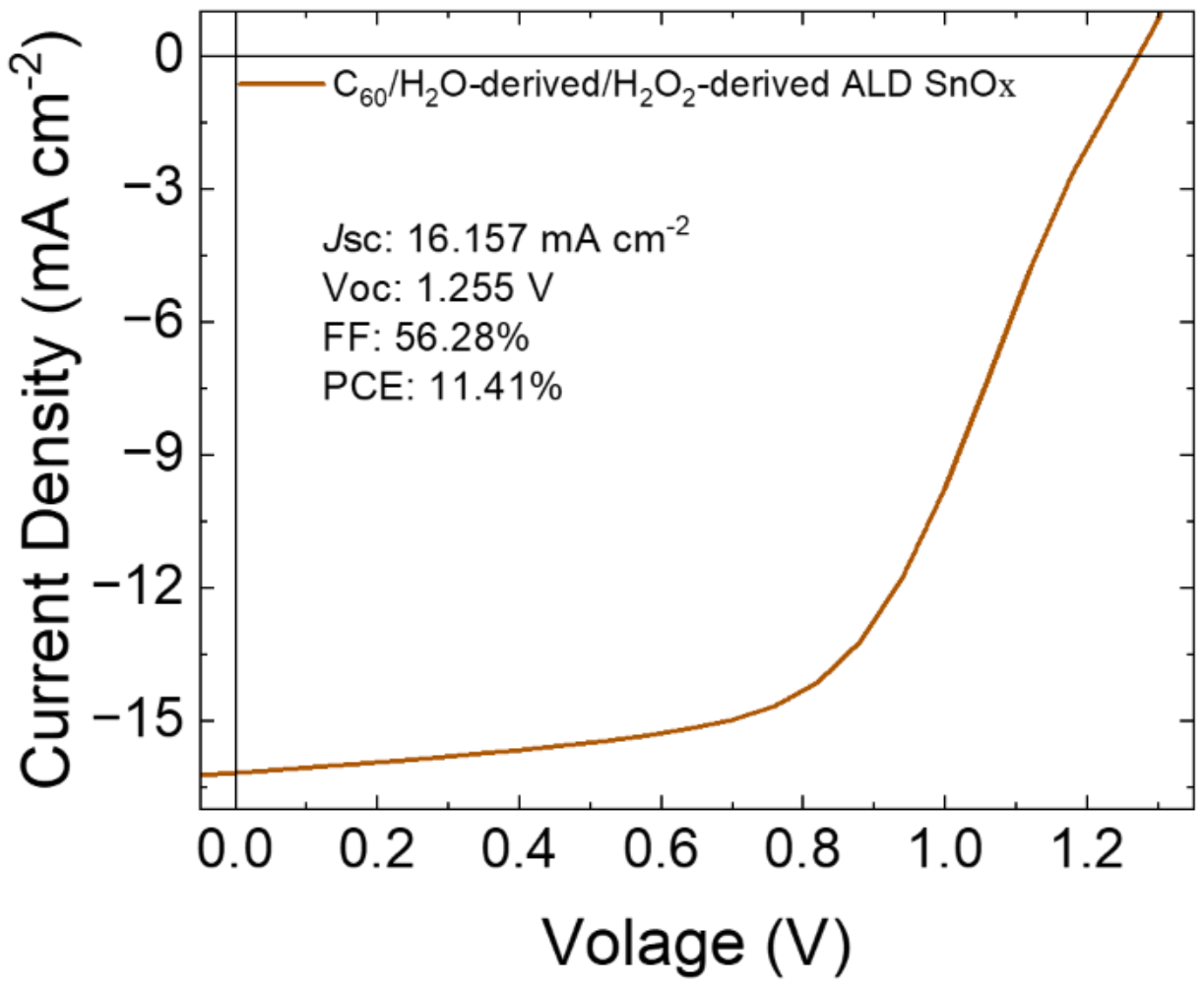

**Figure S6.** The *J-V* curve of P-Q tandem device combining with a $C_{60}/H_2O/H_2O_2$-derived $SnO_x$ ICL.

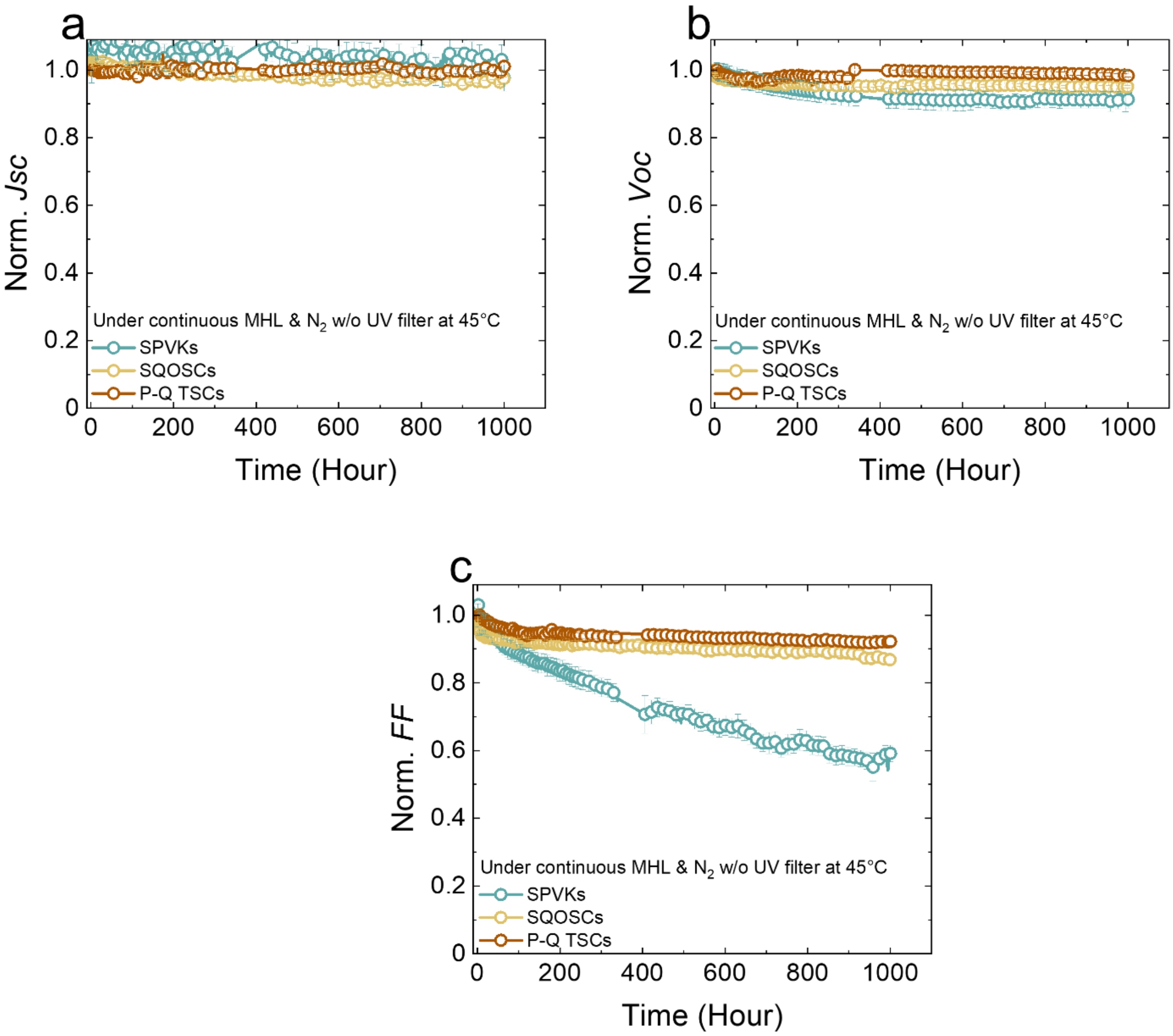

**Figure S7. Evolution of normalized power output of unsealed SOCSs, SPVKs, and P-Q TSCs for 1000 hours of continuous MHL illumination without the UV blocker in a nitrogen flow at short-circuit mode. a**, Norm. *J*sc; **b**, Norm. *V*oc; c, Norm. *FF*.

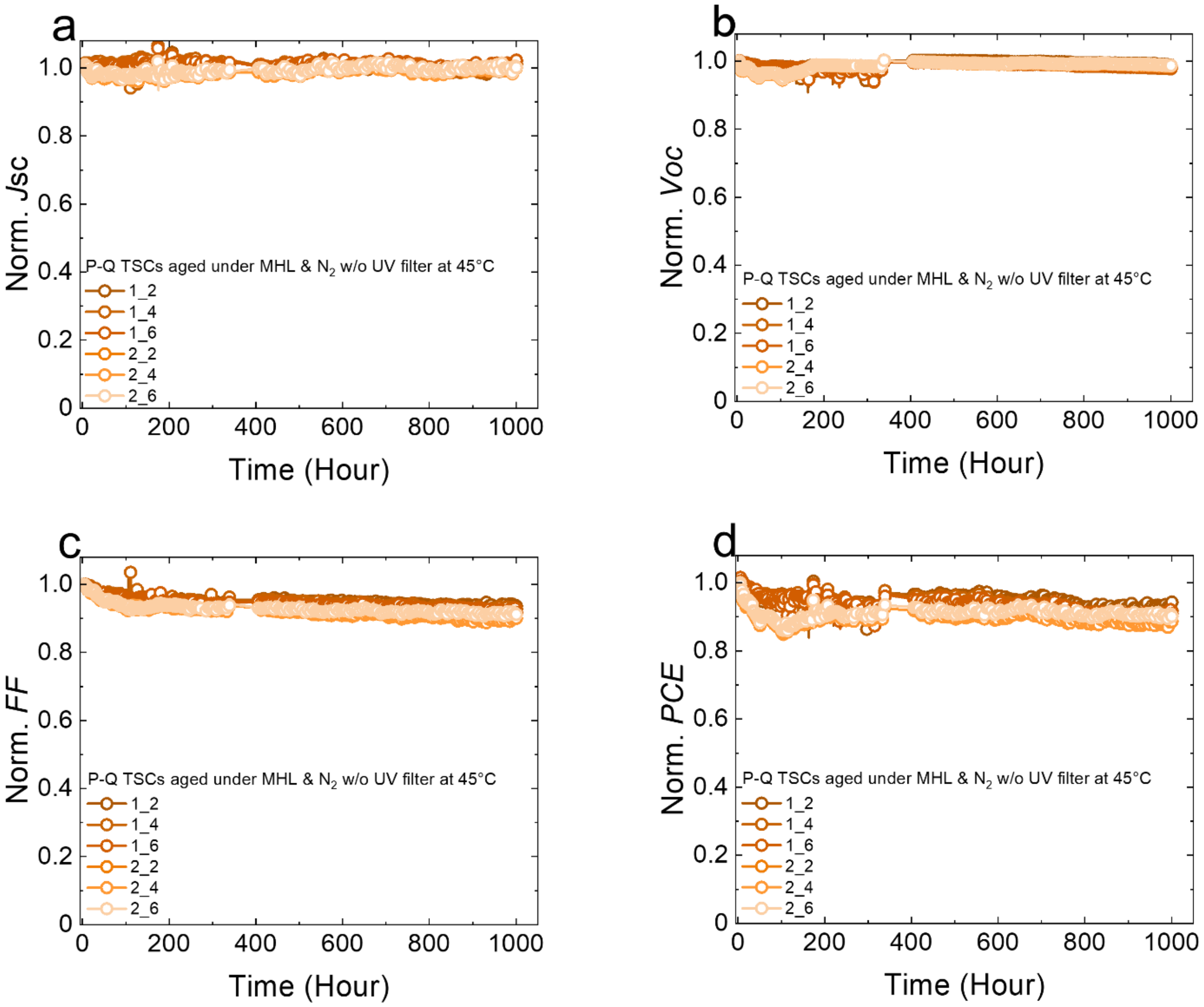

**Figure S8. Stability reproducibility of unsealed P-Q TSCs for 1000 hours of continuous MHL illumination without an extra UV filter in a nitrogen flow at short-circuit mode. a,** Norm. *J*sc; **b**, Norm. *V*oc; c, Norm. *FF*; **d**, Norm. PCE.

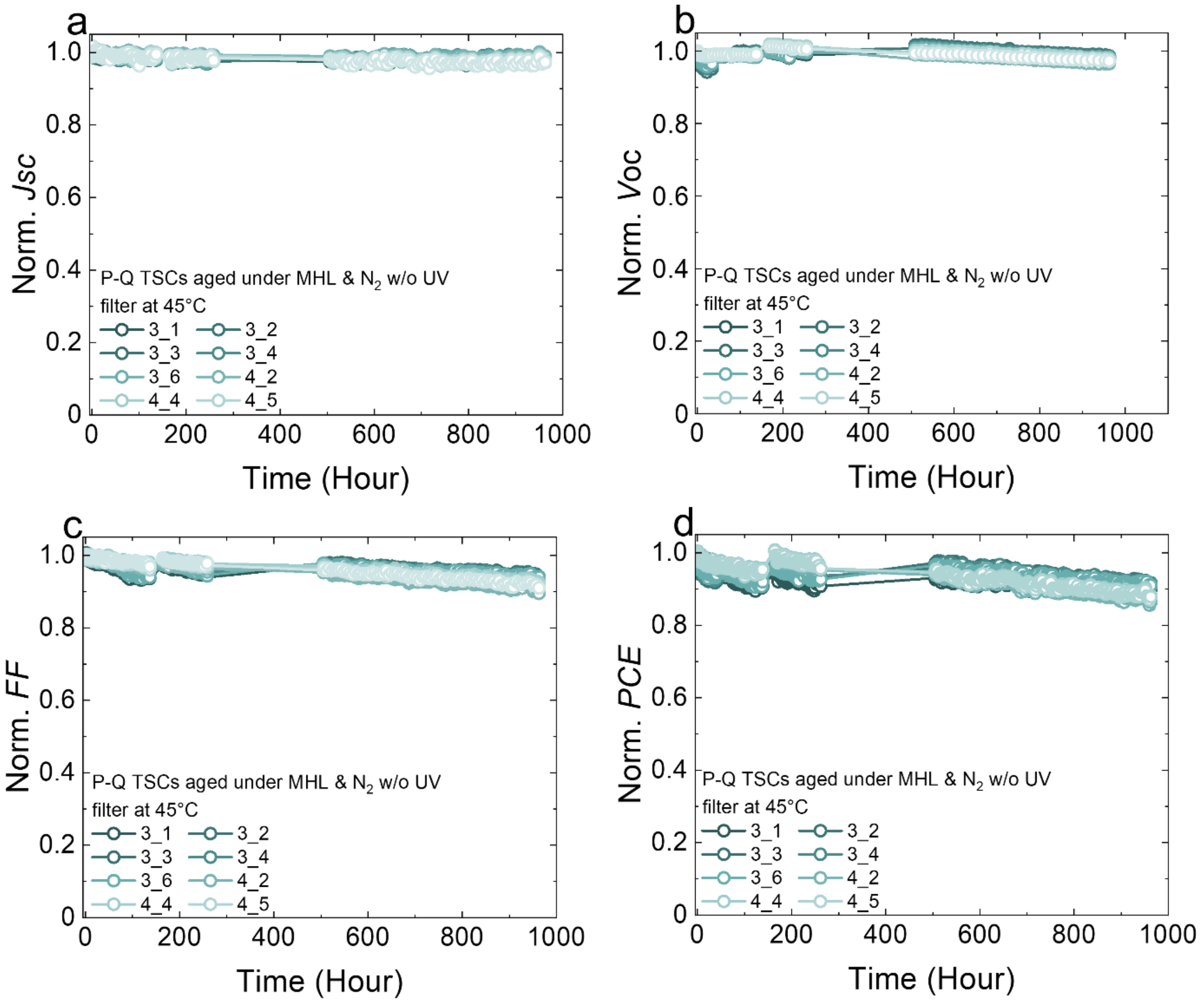

**Figure S9. Stability reproducibility of unsealed P-Q TSCs for near 1000 hours of continuous MHL illumination without an extra UV filter in a nitrogen flow at short-circuit mode. a**, Norm. *J*sc; **b**, Norm. *V*oc; **c**, Norm. *FF*; **d**, Norm. PCE.

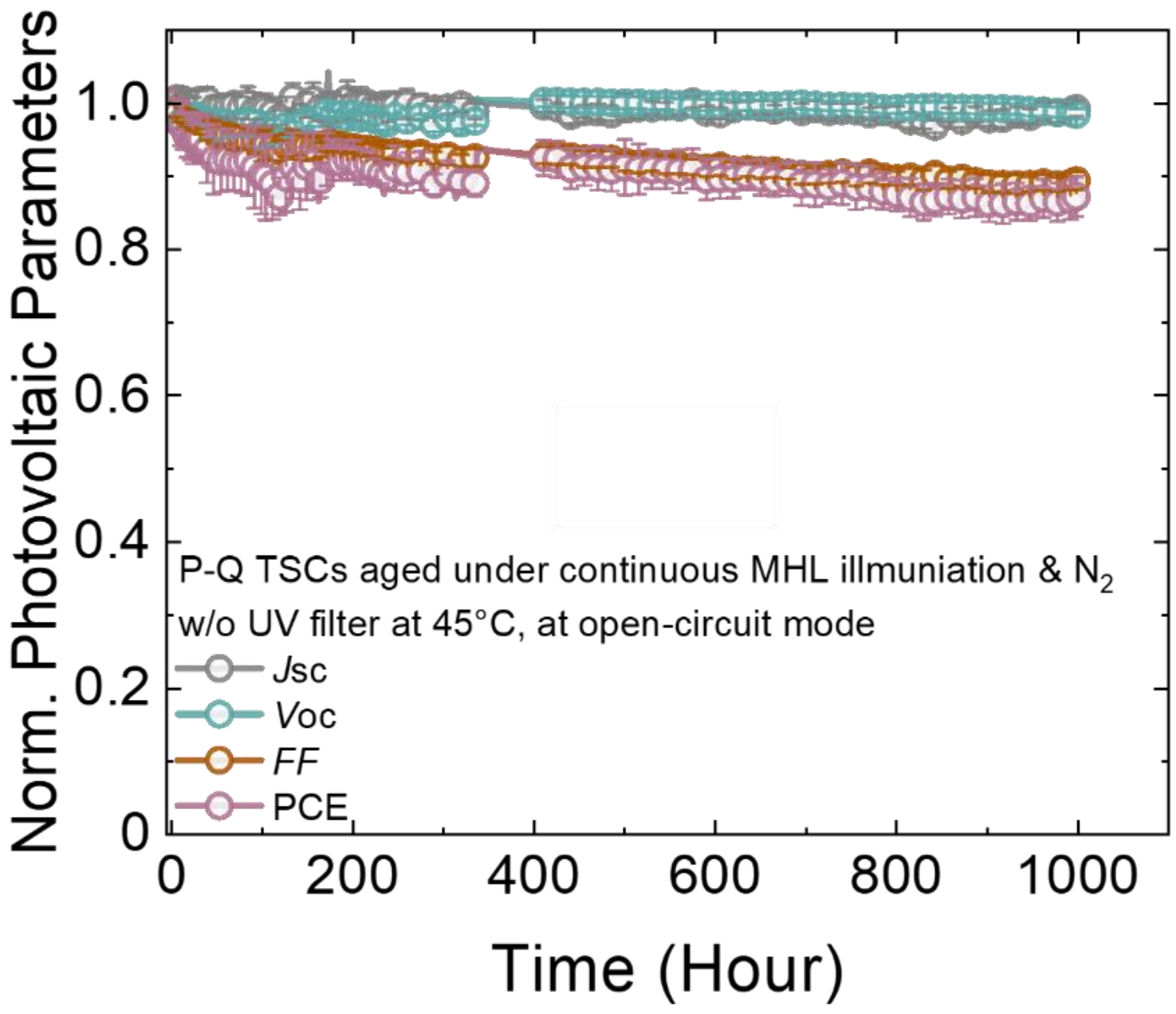

**Figure S10. Evolution of normalized photovoltaic parameters of unsealed P-Q TSCs for 1000 hours of continuous MHL illumination without the UV filter in a nitrogen flow at open-circuit mode.**

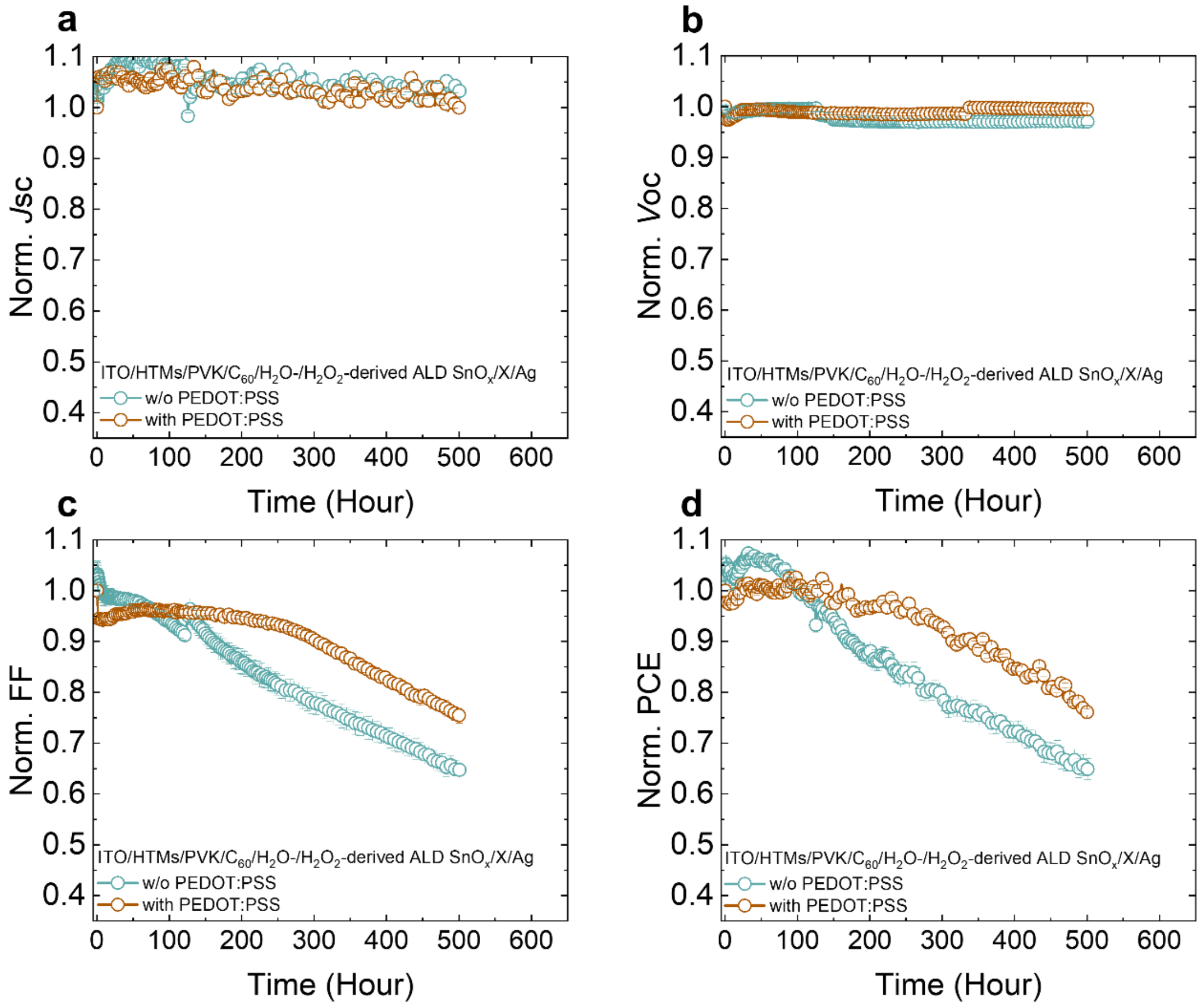

**Figure S11. Stability of the SPVKs with and without PEDOT: PSS layer measured under**

**continuous MHL illumination & $N_2$ of 500 hours without the UV filter under short-circuit mode**. **a**, Norm. Jsc; **b**, Norm. Voc; **c**, Norm. FF; **d**, Norm. PCE.

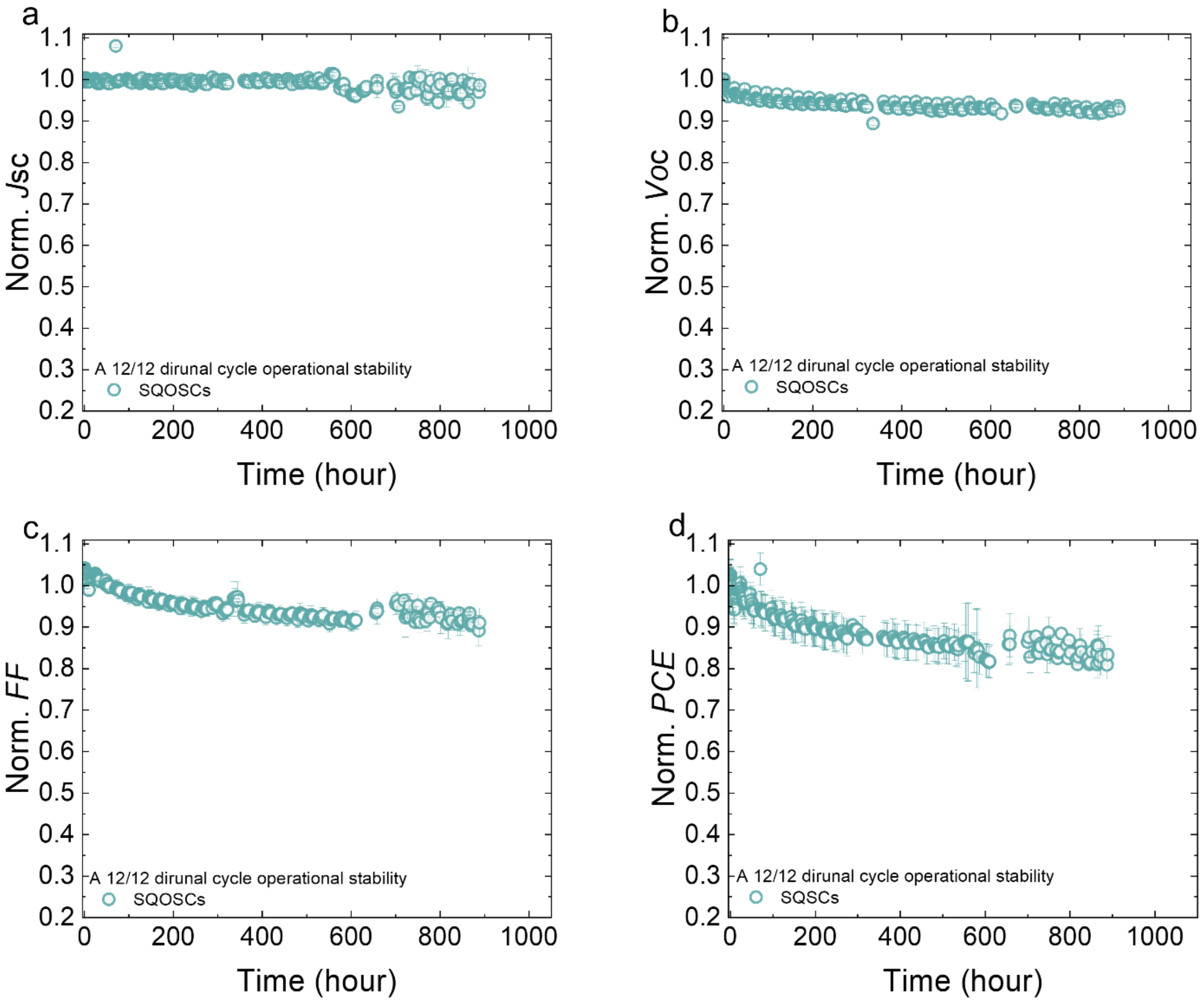

**Figure S12. Evolution of a 12/12 h diurnal cycle operational stability of SQOSCs under MHL illumination without the UV filter in a nitrogen flow & 45 °C at short-circuit mode.** **a**, Norm. *J*sc; **b**, Norm. *V*oc; **c**, Norm. *FF*; **d**, Norm. *PCE*.

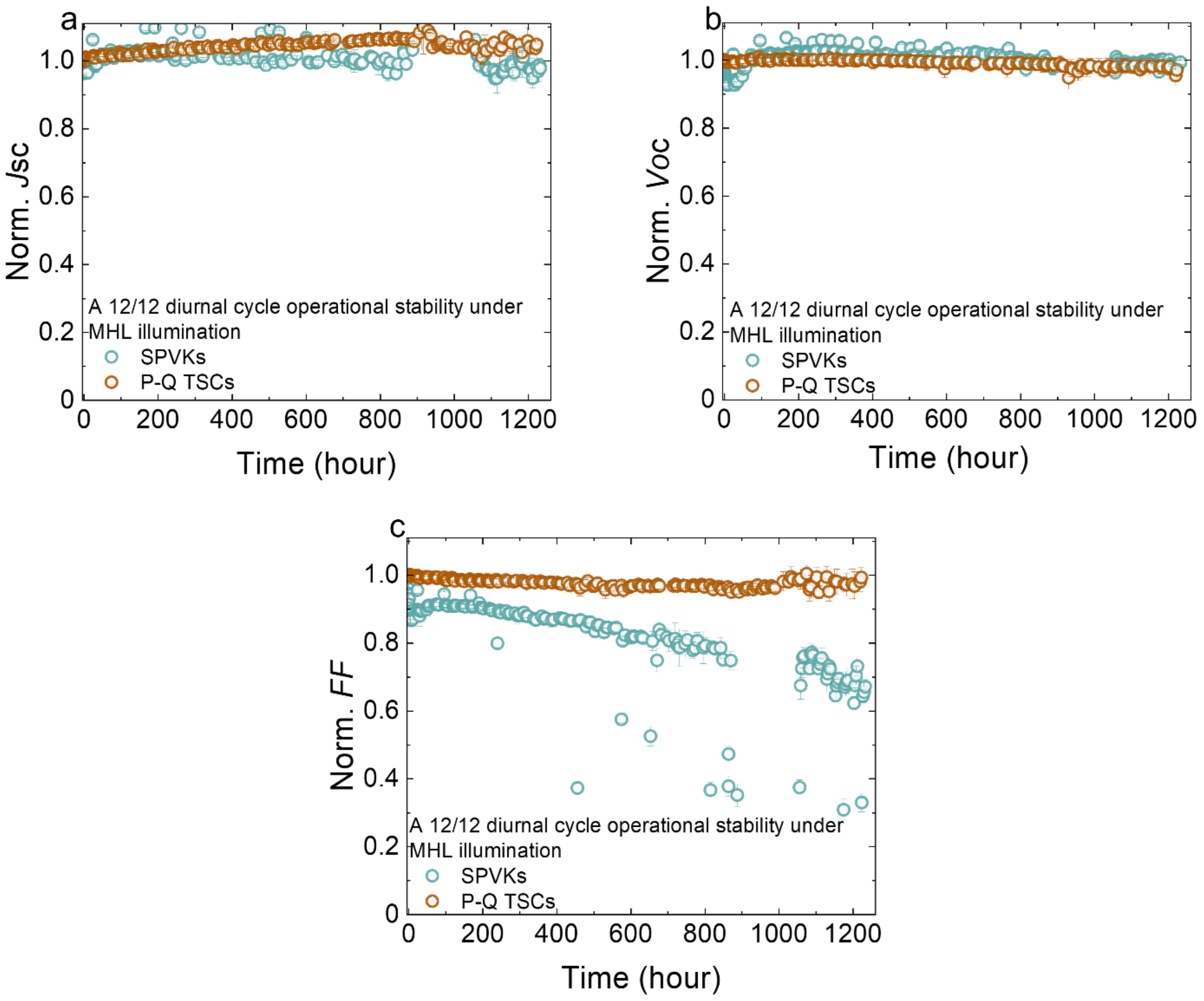

**Figure S13. Evolution of a 12/12 h diurnal cycle operational stability of SPVKs and P-Q TSCs under MHL illumination without an extra UV filter for 61 circles in a nitrogen flow & 45 °C at short-circuit mode. a**, Norm. *J*sc; **b**, Norm. *V*oc; **c**, Norm. *FF*.

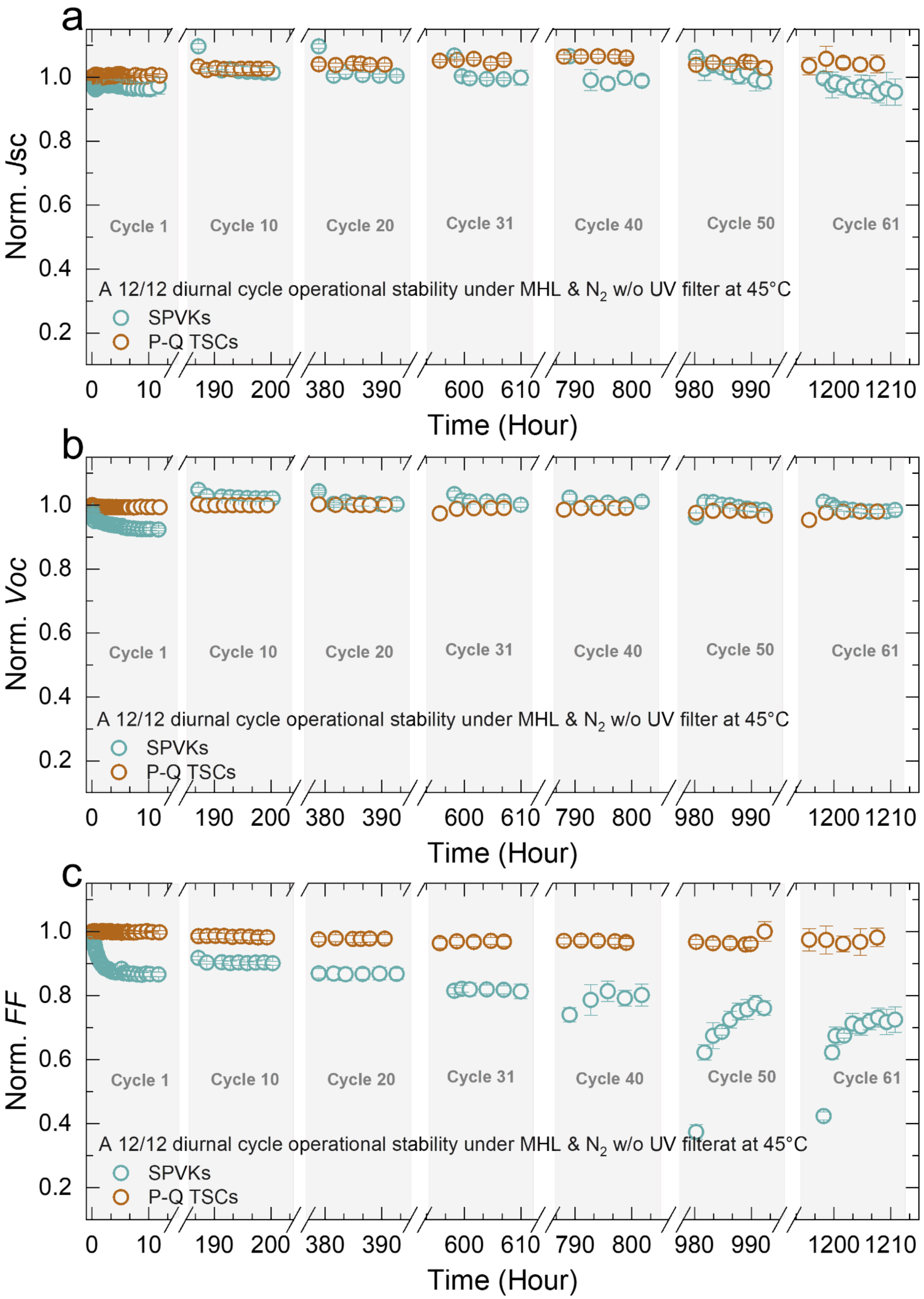

**Figure S14. Evolution of a 12/12 h diurnal cycle operational stability of SPVKs and P-Q TSCs in representative cycles under MHL illumination without the UV filter for 61 circles**

in a nitrogen flow & 45 °C at short-circuit mode. a, Norm. *J*sc; b, Norm. *V*oc; c, Norm. *FF*.

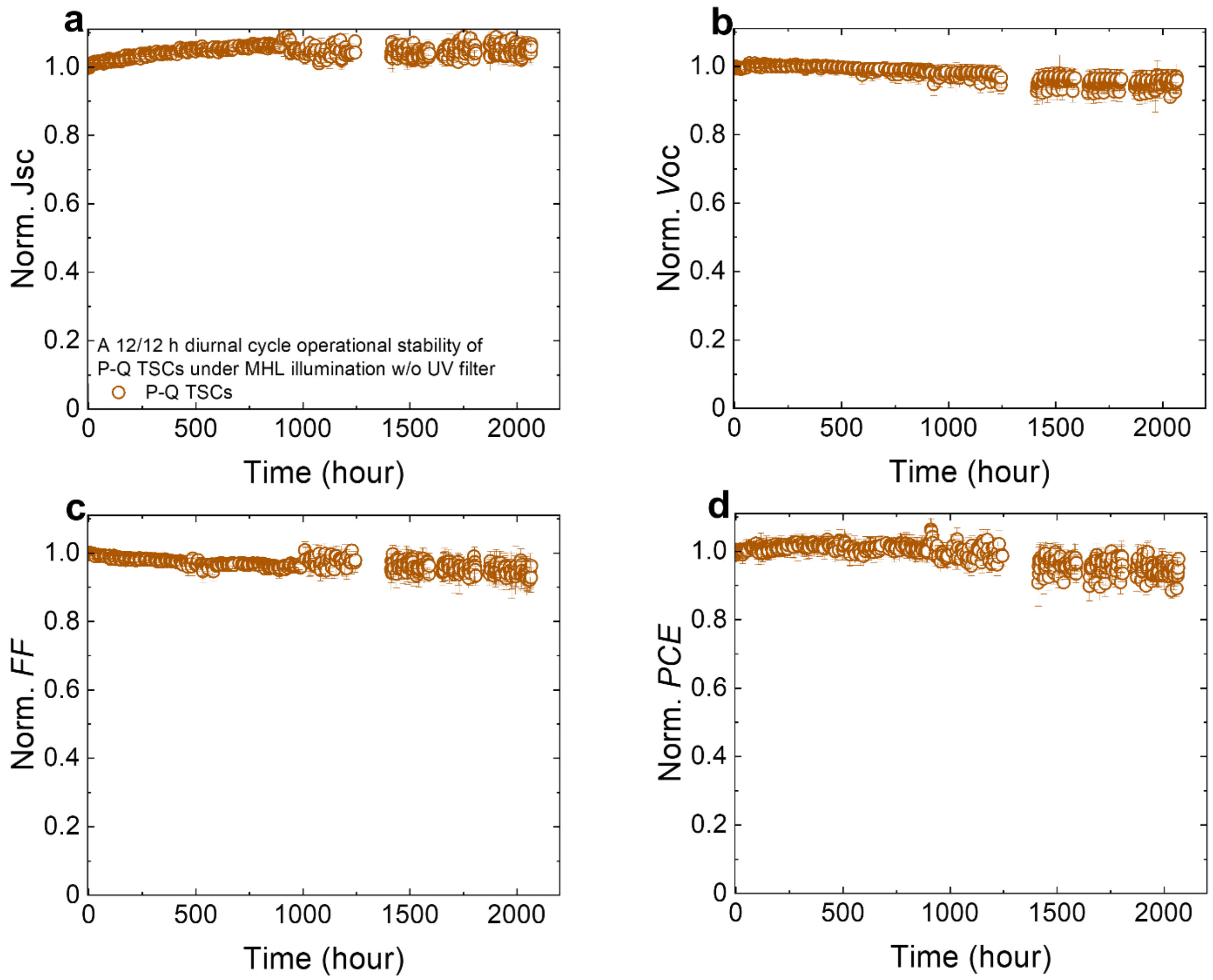

**Figure S15. Evolution of a 12/12 h diurnal cycle operational stability of P-Q TSCs under MHL illumination without the UV filter for 2067 hours (86 circles) in a nitrogen flow & 45 °C at short-circuit mode. a**, Norm. *J*sc; **b**, Norm. *V*oc; **c**, Norm. *FF*; **d**, Norm. *PCE*.

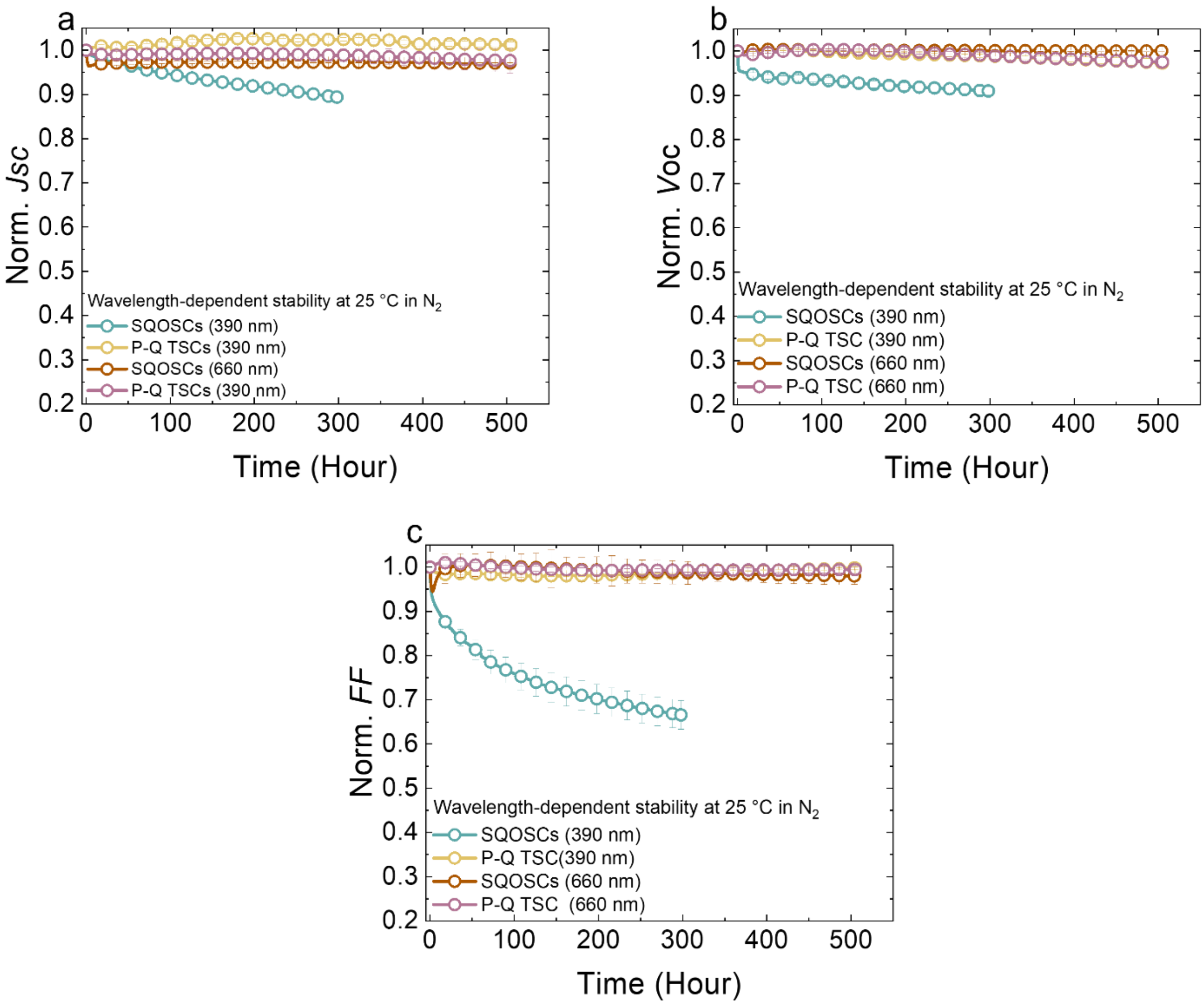

**Figure S16. Evolution of normalized power output of SQOSCs and P-Q TSCs illuminated under 390nm and 660 nm monochromatic sources in a nitrogen flow for 500 hours. a**, Norm. *J*sc; **b**, Norm. *V*oc; **c**, Norm. *FF*.

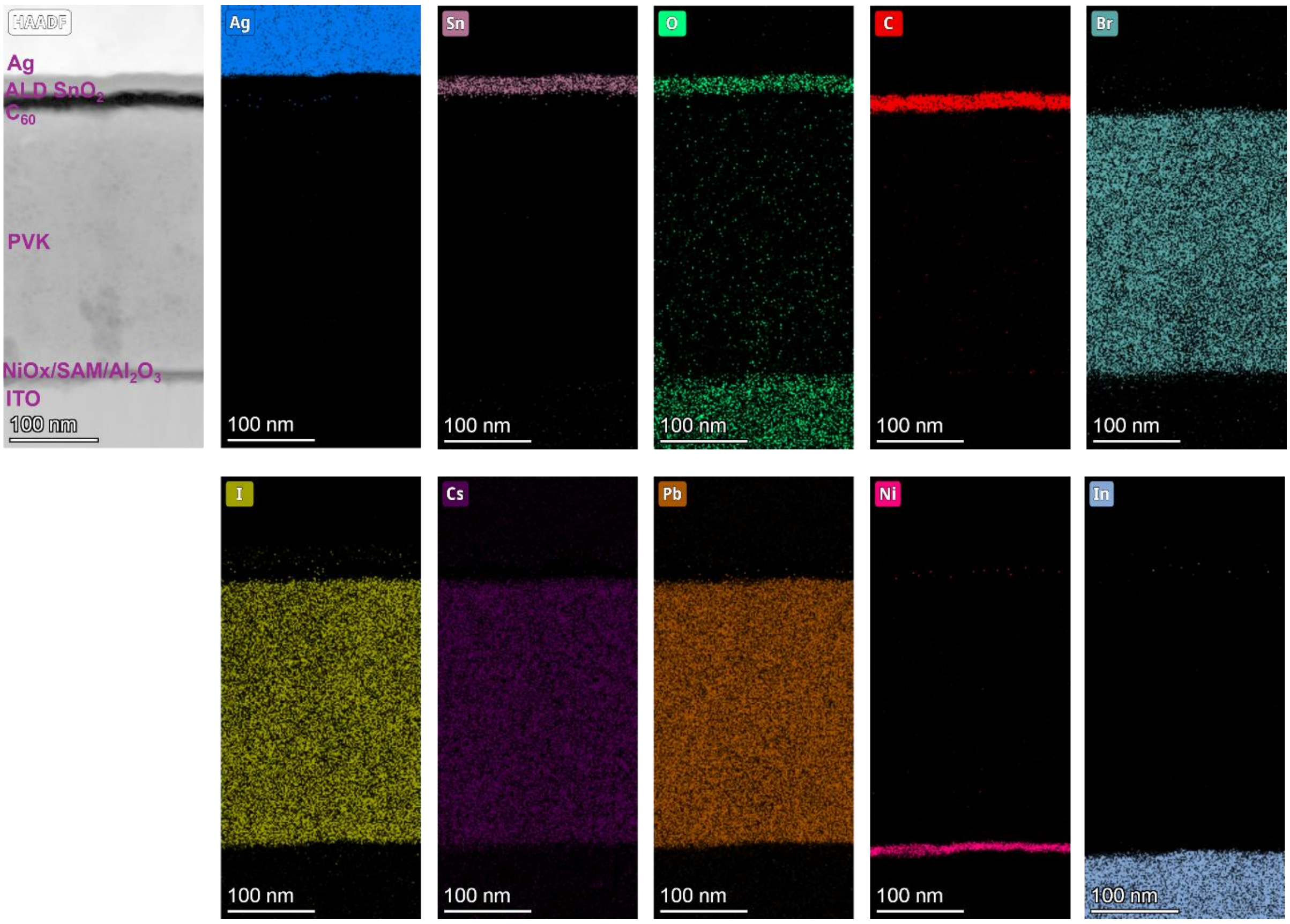

**Figure S17.** Cross-sectional HAADF-STEM image and the corresponding elemental EDXS maps of the fresh SPVK device.

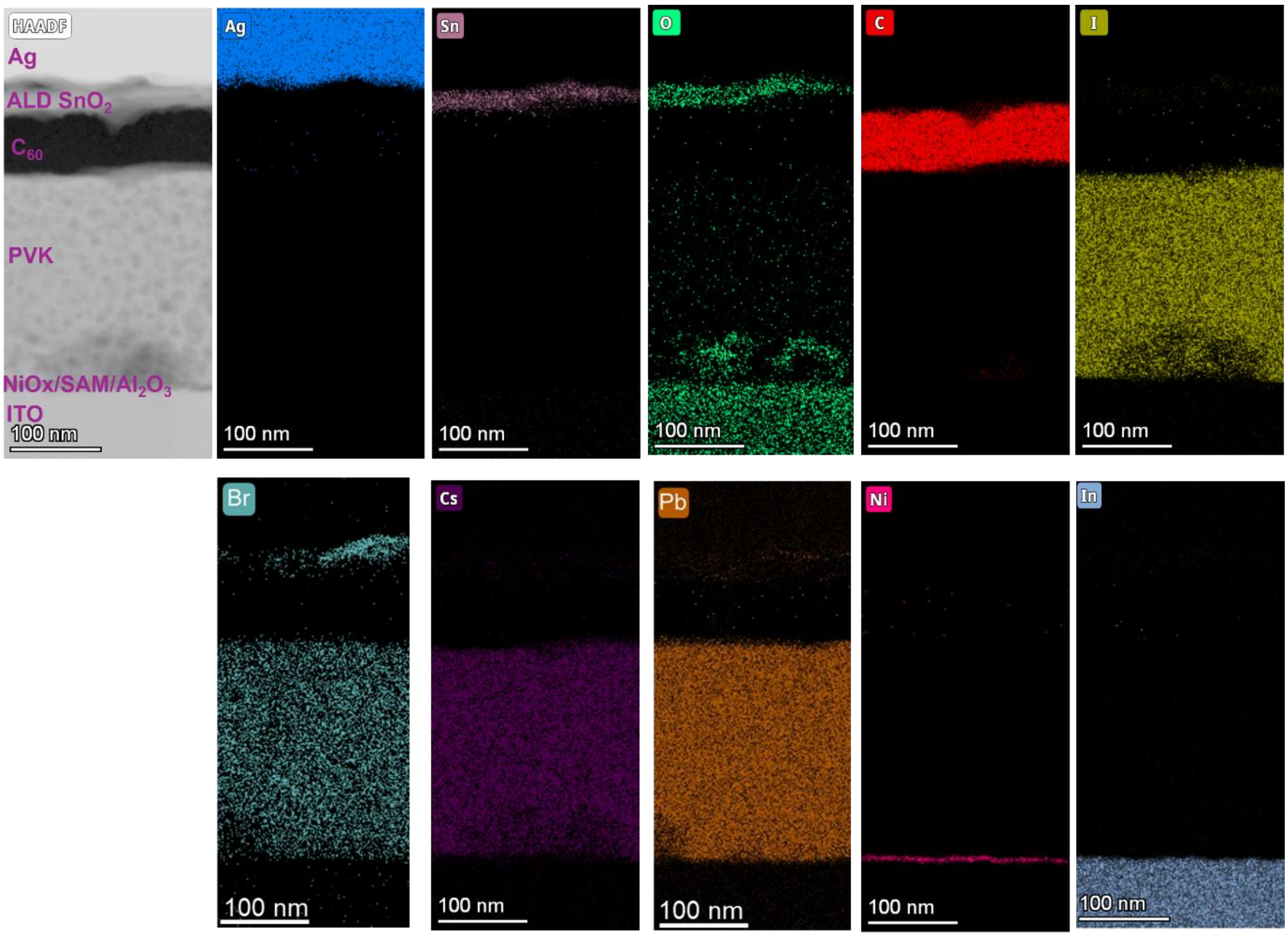

**Figure S18.** Cross-sectional HAADF-STEM and the corresponding elemental EDXS maps of the aged SPVK device.

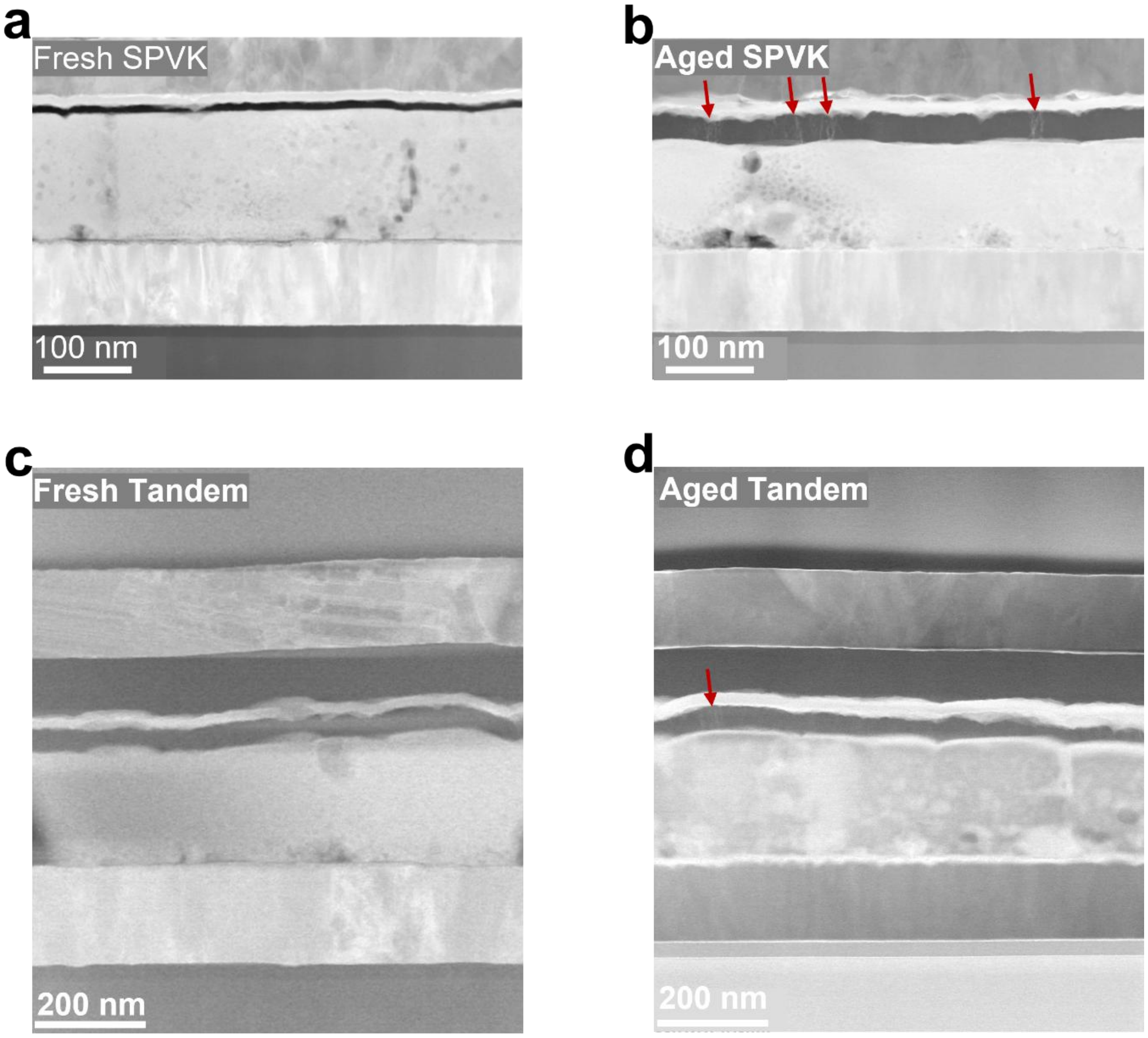

**Figure S19.** HAADF-STEM images of SPVKs and P-Q TSCs. **a, b**, the fresh and aged SPVK, respectively; **c, d**, the fresh and aged P-Q TSC, respectively.

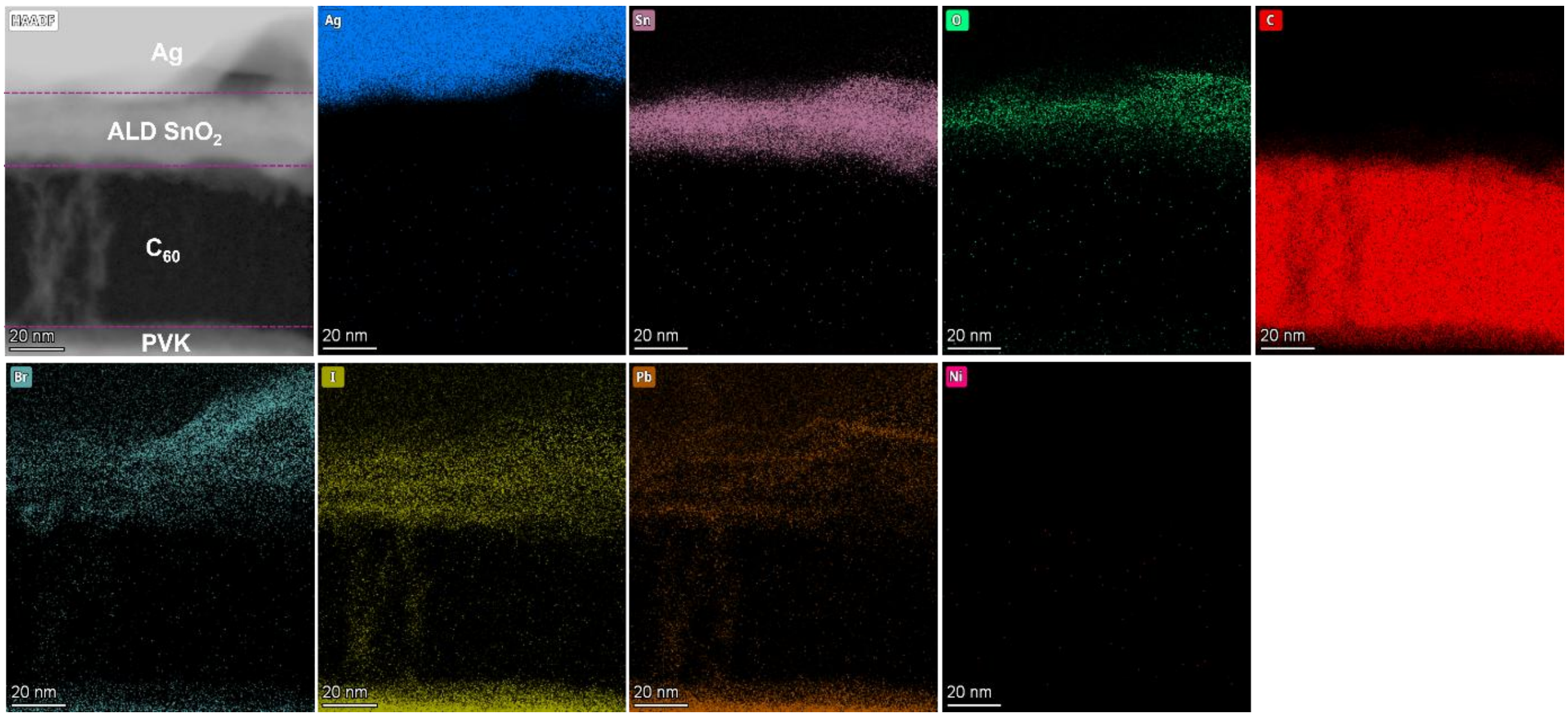

**Figure S20.** Cross-sectional HAADF-STEM image and the corresponding elemental EDXS maps of the aged SPVK.

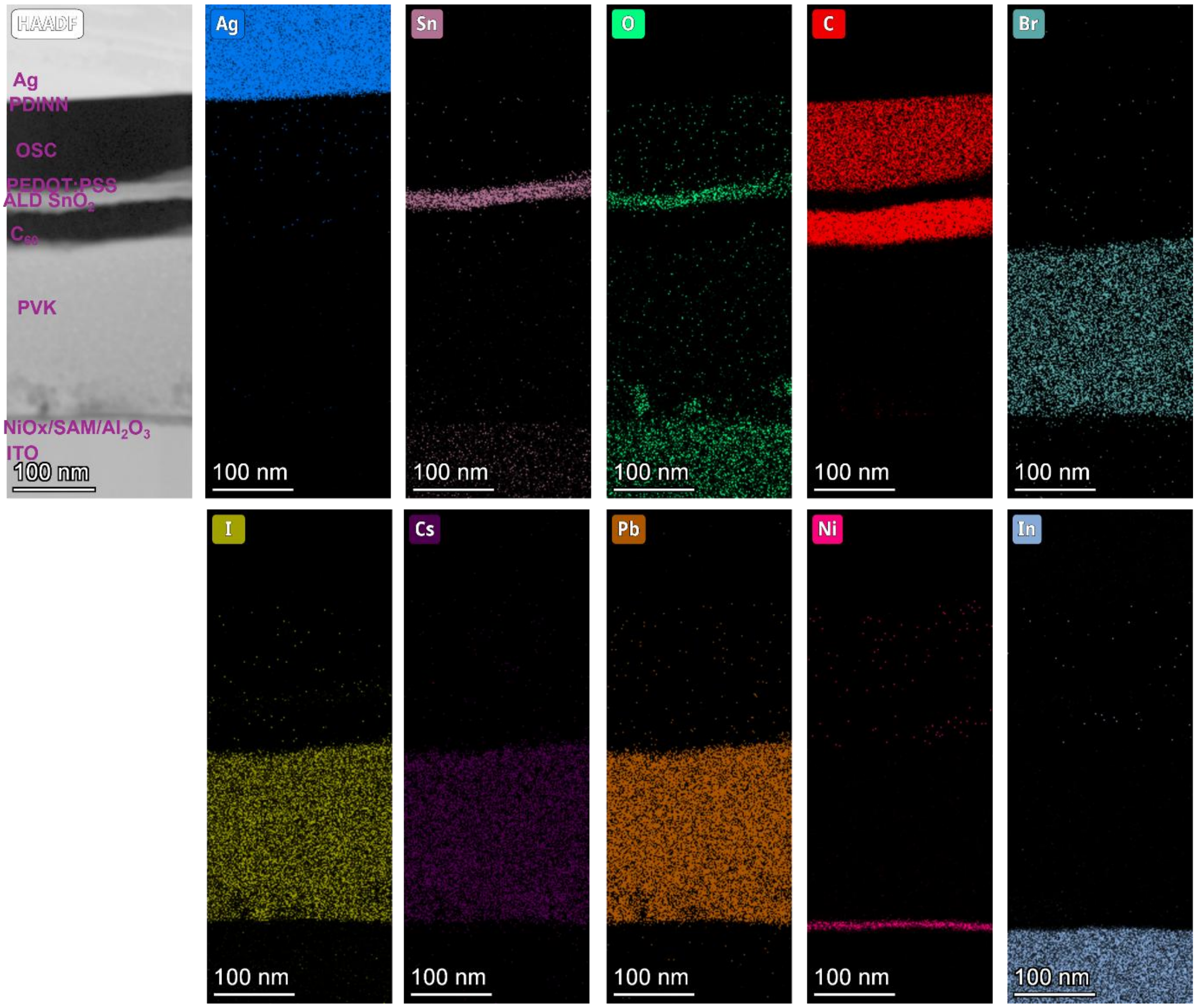

**Figure S21.** Cross-sectional HAADF-STEM image and the corresponding elemental EDXS maps of the fresh P-Q TSC.

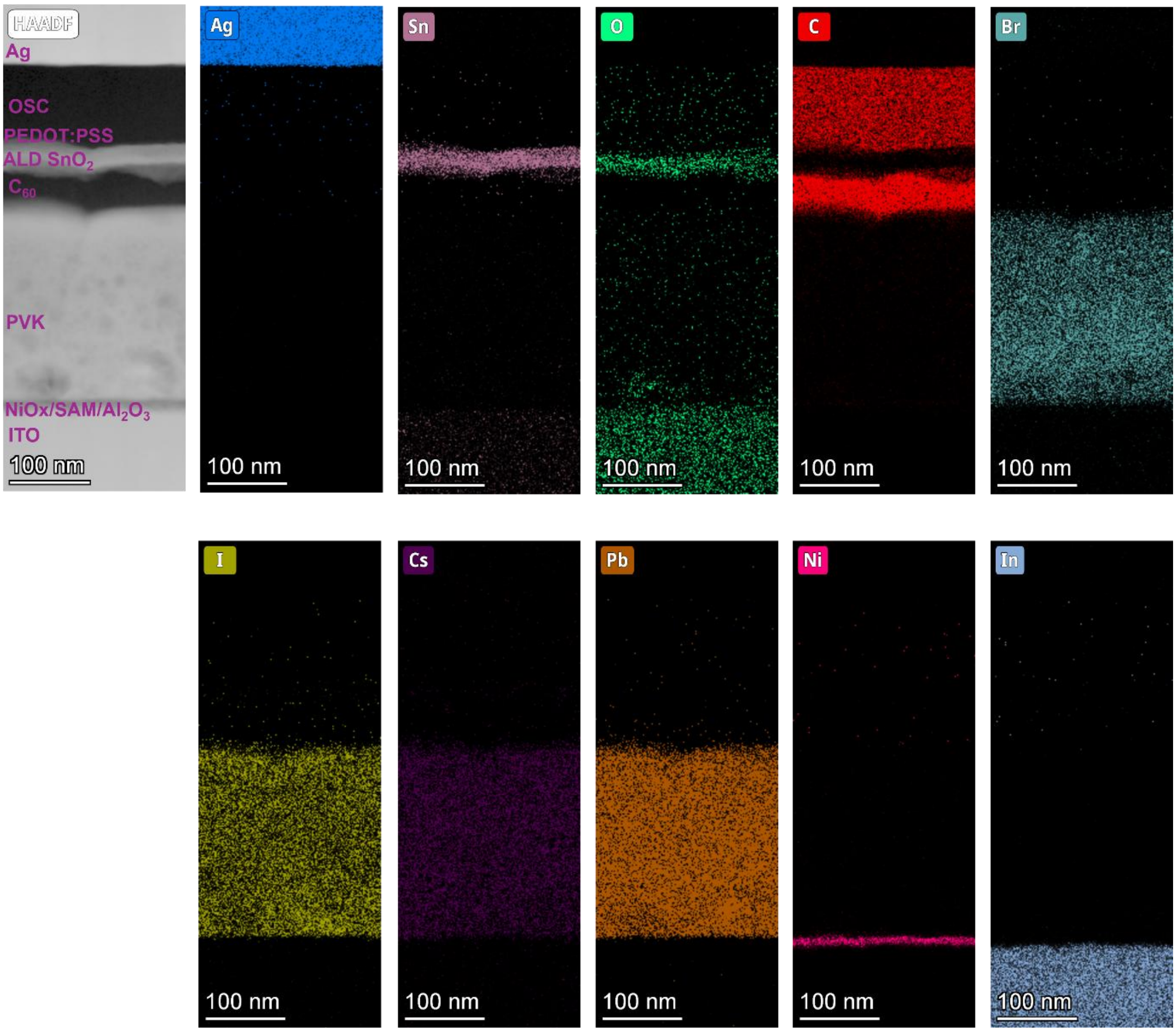

**Figure S22.** Cross-sectional HAADF-STEM image and the corresponding elemental EDXS maps of the aged P-Q TSC.

**Table S1.** Photovoltaic parameters of p-i-n SPVKs with a structure of ITO/$NiO_x$/Me-2PACz/$Al_2O_3$/GABr/1.81 eV perovskite/GABr: F-PEAI/$C_{60}$/various $SnO_x$/Ag. All parameters are measured from a reverse *J-V* scan.

| **Reactant** | **Cycle** | **Scanning direction** | ***Jsc* (mA cm$^{-2}$)** | ***EQE* (mA cm$^{-2}$)** | ***Voc* (V)** | ***FF* (%)** | **PCE$_{\text{-avg.}}$ (%)** |
|---|---|---|---|---|---|---|---|
| $H_2O$ | 140 | Reverse | 16.14±0.19 | 15.76 | 1.25±0.003 | 75.05±0.22 | 15.14±0.26 |
| $H_2O$/ $H_2O_2$ | 70/70 | Reverse | 16.79±0.20 | 16.30 | 1.26±0.01 | 78.66±1.34 | 16.65±0.45 |
| $H_2O$/ $H_2O_2$ | 30/110 | Reverse | 16.60±0.13 | 16.18 | 1.25±0.005 | 77.09±0.01 | 15.99±0.18 |
| $H_2O_2$ | 140 | Reverse | 16.36±0.20 | 15.99 | 1.22±0.003 | 61.88±1.13 | 12.35±0.28 |

**Table S2.** Photovoltaic parameters of p-i-n single-junction OSCs with ternary and quaternary photon-absorbers with a structure of ITO/PEDOT: PSS/organic photon-absorber/PDINN/Ag without anti-refractive film.

| Single-junction OSC | *J*sc (mA cm$^{-2}$) | *EQE* (mA cm$^{-2}$) | *V*oc (V) | *FF* (%) | PCE (%) |
|---|---|---|---|---|---|
| Ternary | 26.28±0.23 | 25.88 | 0.86±0.002 | 76.79±0.58 | 17.36±0.27 |
| Quaternary | 26.71±0.49 | 26.19 | 0.88±0.001 | 77.46±0.54 | 18.21±0.40 |

**Table S3.** Photovoltaic performance of P-T TSCs combining $H_2O$-$SnO_x$ or $H_2O/H_2O_2$-$SnO_x$ based ICLs.

| OSC | ICL | Scanning direction | *J*sc (mA cm$^{-2}$) | *V*oc (V) | *FF* (%) | PCE (%) |
|---|---|---|---|---|---|---|
| Ternary | $H_2O$ | Reverse | -13.96 ±0.18 | 2.07 ±0.006 | 75.26 ±0.85 | 21.75 ±0.37 |
| | | Forward | -14.00 ±0.057 | 2.07 ±0.004 | 74.98 ±0.65 | 21.73 ±0.003 |
| Ternary | $H_2O/H_2O_2$ | Reverse | -14.58 ±0.34 | 2.08 ±0.003 | 77.58 ±0.34 | 23.53 ±0.34 |
| | | Forward | -14.60 ±0.33 | 2.07 ±0.34 | 77.20 ±0.35 | 23.33 ±0.50 |

**Table S4.** Summary of reports on performance of p-i-n P-O TSCs with different bandgap combinations of front PVK cells and rear OPV cells by employing different ICLs.

| Front cell (Eg) | Rear cell (Eg) | ICL | *J*sc (mA cm$^{-2}$) | *V*oc (V) | *FF* (%) | PCE (%) | Reference |
|---|---|---|---|---|---|---|---|
| $Cs_{0.25}FA_{0.75}Pb(Br_{0.5}I_{0.5})_3$ (1.86 eV) | PM6-BTP-eC9 (1.35 eV) | PCBM/AZO/ITO/$MoO_x$ | 14.65 | 2.144 | 80.02 | 25.13 | *Joule* ***8***. 2554 (2024) |
| $FA_{0.8}Cs_{0.2}Pb(I_{0.6}Br_{0.4})_3$ (1.77 eV) | PM6: BTP-eC9:$PC_{71}BM$ (1.35 eV) | $C_{60}$/PEI/ITO/$MoO_x$ | 14.86 | 2.09 | 78.35 | 24.33 | *Adv. Mater.* 2312704 (2024) |
| $FA_{0.8}MA_{0.02}Cs_{0.18}PbI_{1.8}Br_{1.2}$ (1.77 eV) | PM6:Y6:$PC_{71}BM$ (1.36 eV) | $C_{60}$/BCP/Ag/$MoO_x$ | 13.13 | 1.902 | 81.5 | 20.4 | *Joule*, ***4***. 1594-1606 (2020) |
| $MA_{0.96}FA_{0.1}PbI_2Br(SCN)_{0.12}$ (1.72 eV) | PM6:CH1007 (1.36 eV) | PCBM/BCP/Au/$MoO_x$ | 13.8 | 1.96 | 78.4 | 21.2 | *Adv. Funct. Mater.* **32**, 2112126 (2022) |
| $FA_{0.8}Cs_{0.2}PbI_{1.8}Br_{1.2}$ (1.79 eV) | PM6:Y6 (1.36 eV) | $C_{60}$/ALD $SnO_x$/Au/PEDOT:PSS | 13.92 | 2.072 | 77.29 | 22.29 | *Small* ***18***, 2204081 (2022) |
| $FA_{0.6}MA_{0.4}Pb(I_{0.6}Br_{0.4})_3$ (1.78 eV) | PTB7-Th: BTPV-4CL-eC9 (1.22 eV) | $C_{60}$/BCP/Ag/$MoO_x$ | 15.7 | 1.88 | 74.6 | 22 | *Adv. Mater.* ***34***, 2108829 (2022) |
| $FA_{0.8}Cs_{0.2}Pb(I_{0.5}Br_{0.5})_3$ (1.85) | PM7:PM6:Y6 : $PC_{71}BM$ (1.36 eV) | $C_{60}$/BCP/Au/$MoO_x$ | 14.17 | 2.14 | 80.71 | 24.47 | *Adv. Mater*. **35**, 2305946 (2023) |
| $(FAMACsPb(I_{0.5}Br_{0.5})_3)$ (1.85 eV) | PM6:BTP-eC9:$PC_{71}BM$ (1.35 eV) | $C_{60}$/$SnO_2$/Au/$MoO_x$ | 14.15 | 2.197 | 77.6 | 24.12 | *Adv. Mater.* ***36***, *2306568 (2024)* |
| $FA_{0.8}Cs_{0.2}PbI_{1.6}Br_{1.4}$ (1.83 eV) | D18-Cl:N3:$PC_{61}BM$ (1.38 eV) | $C_{60}$/BCP/Ag/$MoO_x$ | 14.36 | 2.11 | 79.38 | 24.05 | *Chin. J. Chem.* **42**, 1819-1827 (2024) |
| $Cs_{0.2}FA_{0.8}Pb(I_{0.6}Br_{0.4})_3$ (1.81 eV) | PM6:Y6: $PC_{71}BM$ (1.36 eV) | $C_{60}$/BCP/Au/$MoO_x$ | 14.36 | 2.15 | 81.56 | 25.22 | *Nat. Energy* ***9***, 4 11-421 (2024) |
| $FA_{0.8}Cs_{0.2}PbI_{1.6}Br_{1.4}$ (1.83 eV) | D18-Cl:N3:$PC_{61}$BM (1.38 eV) | $C_{60}$/BCP/Ag/$MoO_x$/ 2PACZ | 14.68 | 2.12 | 82.35 | 25.82 | *Nat. Energy* ***9***, 592-601 (2024) |
| $FA_{0.8}Cs_{0.2}Pb(I_{0.5}Br_{0.5})_3$ (1.85 eV) | PM6:Y6: $PC_{71}$BM (1.36 eV) | SnOx/$InO_x$/ $MoO_x$ | 14 | 2.15 | 80 | 24 | *Nature* **604**, 280-286 (2022) |
| $Cs_{0.25}FA_{0.75}Pb(I_{0.6}Br_{0.4})_3$ (1.79 eV) | PM6:Y6 (1.36 eV) | $C_{60}$/BCP/IZO/$MoO_x$ | 14.83 | 2.063 | 77.6 | 23.6 | *Nat. Energy 7*, 229-237 (2022) |
| $FA_{0.8}Cs_{0.2}$ Pb $(I_{0.6}Br_{0.4})_3$ (1.77 eV) | PM6:BTP-eC9:$PC_{71}$BM (1.35 eV) | C-C1-P/ITO/$MoO_x$ | 14.58 | 2.09 | 78.99 | 24.07 | *Adv. Mater.* **35**, 2307502 (2023) |
| $FA_{0.8}Cs_{0.2}Pb(Br_{0.4}I_{0.6})_3$ (1.78 eV) | PTB7-Th:IEICO-4F (1.24 eV) | $C_{60}$/ALD SnOx/PEDOT:PSS | 13.8 | 1.83 | 69.4 | 17.6 | *ACS Appl. Energy Mater. **5**, 14035–14044 (*2022*)* |
| $FA_{0.6}MA_{0.4}Pb(I_{0.6}Br_{0.4})_3$ (1.77 eV) | PM6:Y6 (1.36 eV) | PCBM/SnOx/PEDOT:PSS | 14.08 | 2.12 | 74.95 | 22.31 | *Adv. Funct. Mater. **33**, 2308794 (*2023*)* |
| FA0.70MA0.20Rb0.10Pb(I0.5Br0.5)3 (1.88 eV) | PM6:BTPSe-Ph4F (1.27 eV) | $C_{60}$/$SnO_x$/Au/PEDOT:PSS | 15.4 | 2.16 | 79.4 | 26.4 | *Nature* **635**, 860-866 (2024) |
| Cs0.25FA0.75 Pb(Br0.4I0.6)$_3$ (1.8 eV) | PM6:P2EH-1V (1.27 eV) | $C_{60}$/SnO*x*/ITO/MoO*x* | 15.37 | 2.14 | 83.7 | 27.5 | *Nature* **643**,104-110 (2025) |
| $Cs_{0.3}FA_{0.7}Pb(I_{0.6}Br_{0.4})_3$ (1.81 eV) | PM6: L8BO: BTP-eC9 :$PC_{70}$BM (1.4 eV) | $C_{60}$/ALD SnOx/PEDOT:PSS | 15.38 | 2.11 | 77.40 | 25.12 | *This work* |